%% file: topcouplings.tex
\documentclass[10pt]{article}
\usepackage{jheppub}
\usepackage{graphicx}
\usepackage{float}
\usepackage{amsmath}
\usepackage{amsfonts}
\usepackage{subfigure}
\usepackage{multirow}
\usepackage{colortbl}
\definecolor{kugray5}{RGB}{224,224,224}
\usepackage{color}
\input macro.tex

\def\home{.}
\title{Top couplings}

\author{
Jahred Adelman,\\
Department of Physics, Yale University, New Haven, CT 06511\\
\vskip 0.4cm
Barbara Alvarez Gonzalez,\\
Department of Physics and Astronomy, Michigan State University, East Lansing MI 48824\\
\vskip 0.4cm
Yang Bai,\\
Department of Physics, University of Wisconsin, Madison, WI 53706
\vskip 0.4cm
Matthew Baumgart,\\
Department of Physics, Carnegie Mellon University, Pittsburgh, PA 15213\\
\vskip 0.4cm
R. Keith Ellis,\\
Theoretical Physics Department, Fermilab, Batavia, IL 60510\\
\vskip 0.4cm
A. Khanov,\\
Department of Physics, Oklahoma State University, Stillwater OK 74078\\
\vskip 0.4cm
Andrey Loginov,\\
Department of Physics, Yale University, New Haven, CT 06511\\
\vskip 0.4cm
Marcel Vos, \\
IFIC (UVEG/CSIC), Ap. Correos 22085, E46071 Valencia, Spain\\
}

\abstract{Overview on top couplings measurements is presented, and the
  prospects of future measurements are discussed. The coupling of top
  to the W boson can be examined either by looking at the decay of the
  top quark or from single top quark production. With the advent of
  high statistics top physics at the LHC and at the high-luminosity
  LHC, the processes where the bosons (photon, Z and Higgs) are
  produced in association with top quarks become accessible. The first
  evidence on the coupling of the top quark to these particles will
  come from the production rate.}

\keywords{QCD, Phenomenological Models, Hadronic Colliders, LHC}

\begin{document}
%\pagewiselinenumbers
\maketitle

\input BBE/bbe.tex

\section{Top Quark Weak Interaction Measurements\protect\footnote{Author: Barbara Alvarez Gonzalez, Michigan State University}}
\label{s:singletop}
\input Alvarez/Alvarez.tex
\section{Coupling of the Top quark to charge zero vector bosons\protect\footnote{Author: Andrey Loginov, Yale University}}
\input Loginov/symbols.tex
\input Loginov/loginov.tex
\input Vos/vos.tex

\section{Coupling of the Top quark to the Higgs boson\protect\footnote{Author: Jahred Adelman, Yale University}}
\input Adelman/adelman.tex
\section{Top quark + jets\protect\footnote{Author: Sasha Khanov, OK State University}}
\input Khanov/khanov.tex

\vskip 0.3cm
\noindent
{\bf \large Summary} 

Overview on top couplings measurements is presented, and the prospects
of future measurements are discussed. With the advent of high
statistics top physics at the LHC with 300 $fb^{-1}$ and at the
high-luminosity LHC with 3000 $fb^{-1}$, the processes where the
bosons (photon, Z and Higgs) are produced in association with top
quarks become accessible. The first evidence on the coupling of the
top quark to these particles will come from the production rate,
followed by precision measurements.

\vskip 0.3cm
\noindent
{\bf \large Acknowledgments} 

The research of RKE is supported by the US DOE under contract
DE-AC02-06CH11357.  MB is supported by DOE grant DE-FG-03-91ER40682.
    
\bibliography{topcouplings}
\bibliographystyle{JHEP}

\end{document}

%% file: macro.tex
\def\TeV{\ifmmode {\mathrm{\ Te\kern -0.1em V}}\else
                   \textrm{Te\kern -0.1em V}\fi}%
\def\GeV{\ifmmode {\mathrm{\ Ge\kern -0.1em V}}\else
                   \textrm{Ge\kern -0.1em V}\fi}%
\let\tev=\TeV
\let\gev=\GeV

\def\spaa#1.#2.#3{\langle\mskip-1mu{#1}|#2|{#3}\mskip-1mu\rangle}
\def\spbb#1.#2.#3{[\mskip-1mu{#1}|#2|{#3}\mskip-1mu]}
\def\spa#1.#2{\left\langle#1\,#2\right\rangle}
\def\spb#1.#2{\left[#1\,#2\right]}
\def\spab#1.#2.#3{\left\langle#1|#2|#3\right]}
\def\spba#1.#2.#3{\left[#1|#2|#3\right\rangle}

\def\P3b{\bar{P}_3}

\def\g0{\gamma_0}

\newcommand{\beq}{\begin{equation}}
\newcommand{\eeq}{\end{equation}}
\newcommand{\beqn}{\begin{eqnarray}}
\newcommand{\eeqn}{\end{eqnarray}}

\newcommand{\Et}{E_t}
\def\vec#1{{\mbox{\boldmath$#1$}}}

%% file: BBE/bbe.tex
\section{Standard Model couplings\protect\footnote{Author: R. Keith Ellis, Fermilab}}
\begin{figure}[h]
\begin{center}
\includegraphics[angle=270,scale=0.5]{\home/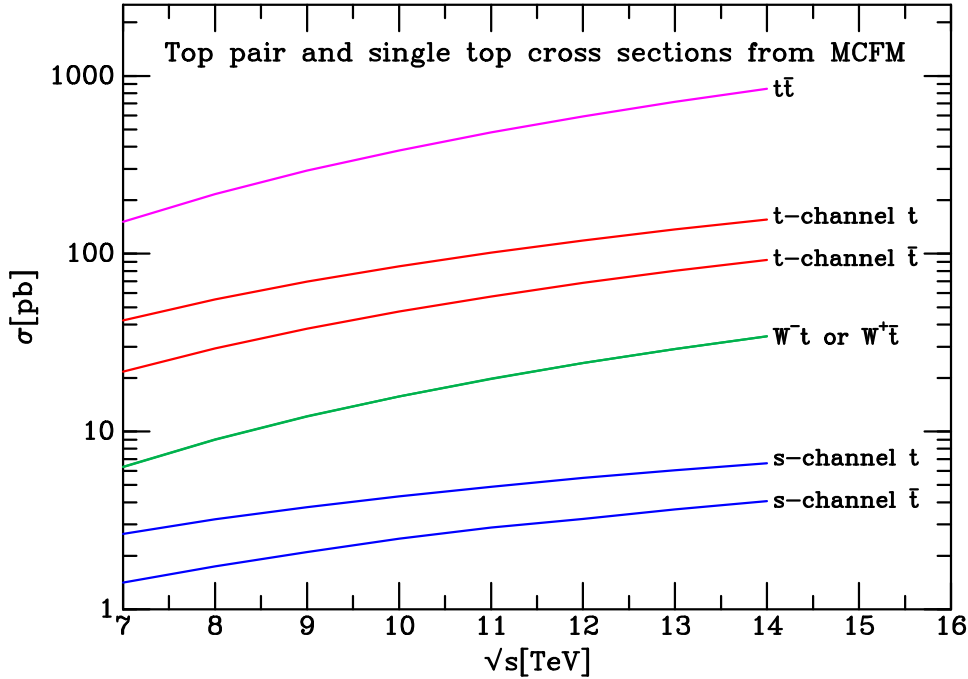}
\caption{NLO values for the top cross sections vs $\sqrt{s}$
calculated using MCFM. The cross sections are evaluated at 
factorization and renormalization scale $m_t=173.2$~GeV using
the CTEQ6M parton distributions.} 
\label{xsec}
\end{center}
\end{figure}

The top quark couples to other Standard Model fields through its gauge
and Yukawa interactions.  At LHC energies the top quark is copiously produced
both in pair production and in the three single top processes mediated by
the exchange or production of a $W$-boson. Fig.~\ref{xsec} shows the cross sections
for these processes as a function of the centre of mass energy. 

With the advent of high statistics top
physics at the LHC, the processes where the bosons, $\gamma,Z$
and $H$ are produced in association with top quarks become accessible.
The first evidence on the coupling of the top quark to these particles
will come from the production rate.  The coupling of top to the
$W$ boson can be examined either by looking at the decay of the top
quark or from single top quark production.  Two recent reviews of the
experimental situation for top couplings can be found in
ref.~\cite{Schilling:2012dx,Liss:2012sm}.

\subsection{Top + $\gamma$}
Experimental results on the production of a photon in association with
a top pair have been presented by both the CDF
collaboration\cite{Aaltonen:2011sp} and the ATLAS
collaboration\cite{2011-153}.  Next-to-leading order (NLO) corrections
to the top cross section in association with a photon are given in
refs.~\cite{Melnikov:2011ta,PengFei:2011qg}. When confronting theory
with experimental data it is important to include photon radiation off
top quark decay products, which are found to give a significant
contribution to the cross-section~\cite{Melnikov:2011ta}.

\begin{figure}
\begin{center}
\includegraphics[angle=270,scale=0.6]{\home/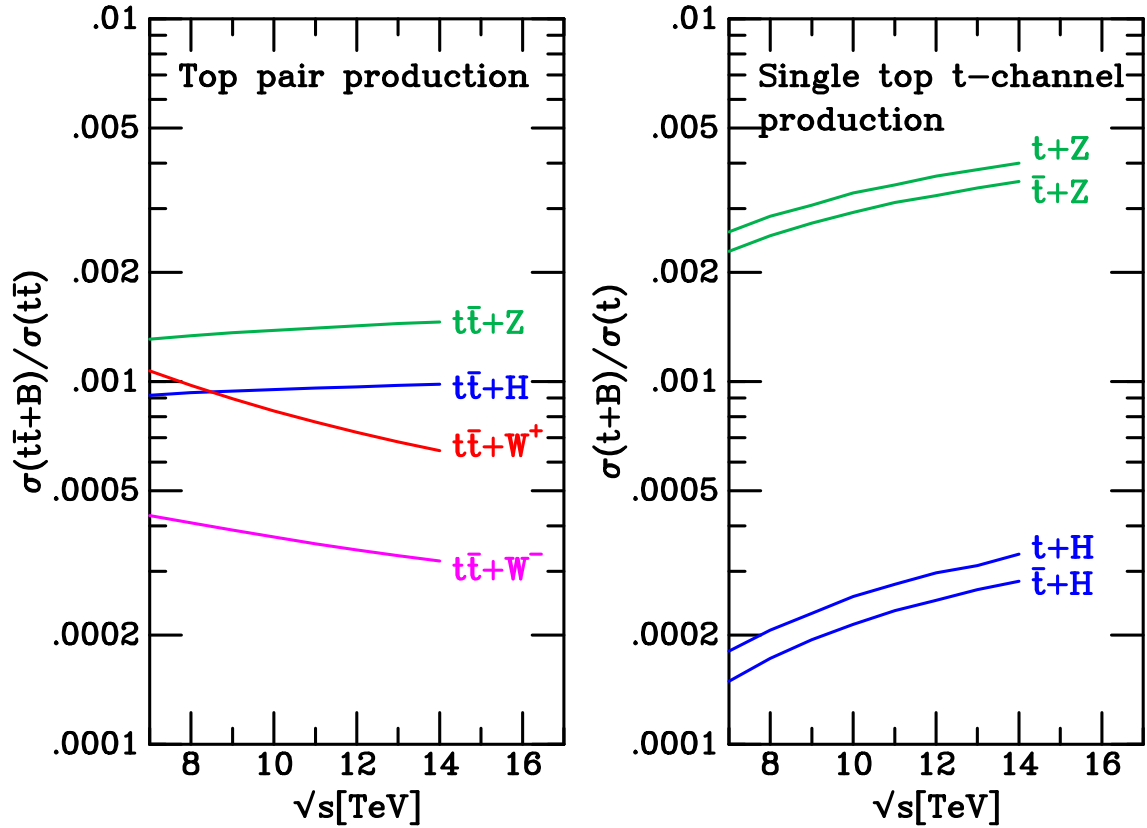}
\caption{Fraction of top pair (left pane) and t-channel single-top production which is 
accompanied by a massive boson. The $t \bar{t}+W^\pm$ process is included for completeness,
even though the $W$ does not couple to the top quark. The ratios are calculated in lowest order
perturbation theory.}
\label{combine}
\end{center}
\end{figure}
\subsection{Top + $Z$}
The primary source of information on the coupling of the top to the $Z$-boson 
will come from the associated production of the $Z$ with a top quark.
A first measurement of vector boson production in association with top-antitop pairs
is given in ref.~\cite{ttv_cms}. A search for this final state has also 
been performed by ATLAS~\cite{ttz_atlas}.
Flavor changing couplings of the top quark to the $Z$ are limited by the 
limits on neutral current decay of the top quark~\cite{Aad:2012ij,:2012sd}.

Also of interest is the associated production of a $Z$-boson in single top production, 
since it is sensitive to the couplings of the $Z$ to both the $W$-boson and the quarks,
including the top.
Fig.~\ref{combine} shows the production of $t \bar{t}+Z$ and the production of 
$t +Z$ and $\bar{t}+Z$ as a fraction of the corresponding cross section without a vector boson.
At $\sqrt{s}=14$~TeV the $t\bar{t}$ pair cross section 
is about $1$~nb and the single top production
cross section for $t$ ($\bar{t}$) is about $150(100)$~pb. 

\subsection{Top + $H$}
Indirect evidence of the coupling of the top to the Higgs boson comes 
from the Higgs boson production rate in $pp$ collisions at the LHC. 
The gluon-gluon fusion production of the Higgs boson
is predicted to proceed predominantly through a top loop.
The decay of the Higgs, $H \to \gamma \gamma$ also proceeds through a top loop (and a $W$-boson) loop
and will provide complementary information.

Further information will have to await the observation of the direct production process, $t\bar{t}+H$.
Fig.~\ref{combine} shows the production of $t \bar{t}+H$ and the production of 
$t +H$ and $\bar{t}+H$ as a fraction of the cross section without the corresponding vector boson.
Theoretical predictions for $t \bar tH$ production at next-to-leading order can be found in 
refs.~\cite{Beenakker:2001rj,Reina:2001bc,Dawson:2003zu,Garzelli:2011vp,Frederix:2011zi}.

\begin{figure}
\begin{center}
\includegraphics[angle=270,scale=0.6]{\home/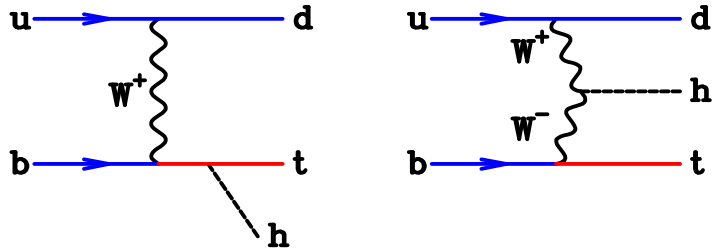}
\caption{Diagrams for the production of the Higgs in association with a single top.}
\label{STHDiags}
\end{center}
\end{figure}

The production of the Higgs in association with a single top is of interest
because of the substantial cancellation between the two diagrams where the Higgs is
emitted from the top quark or from the $W$-boson exchanged in the $t$-channel, as shown in Fig.~\ref{STHDiags}.
After inclusion of the branching ratios the cross section is very small, perhaps beyond 
the limit of observability at the high luminosity LHC.
Thus any non-standard physics that affects the cancellation such as a change of sign of the coupling to the top
will lead to a much larger cross section. 
For a discussion of this process and references to 
recent literature, see refs.~\cite{Maltoni:2001hu,Barger:2009ky,Farina:2012xp,Biswas:2012bd,Agrawal:2012ga}.

\subsection{Top + $W$}

Constraints on the couplings of the top to the $W$ come from top decay and from single top production.
Measurements of the $W$ boson helicity in top decay probes the structure of the 
$Wtb$ vertex. In the standard model the branching fraction of the top quark to longitudinal
$W$ bosons is given by,
\beq
\frac{\Gamma(t\to b W_0)}{\Gamma(t\to b W) }   =F_0= \frac{m_t^2}{m_t^2+2 M_w^2}
\eeq
where we have neglected the mass of the $b$ quark. 
The polarization state of the W controls the angular distribution of the leptons into which it decays. 
We may define the lepton helicity angle $\theta_e^*$, 
as the angle of the charged lepton in the rest frame of the $W$, 
with respect to the original direction of travel of the $W$.
The normalized angular distribution of the charged leptons is therefore
given by, 
\beq
\frac{1}{N }\frac{d N (W\rightarrow e\nu)}{d \cos \theta_e^*}= 
\Bigg[\frac{3}{4} \sin^2 \theta_e^* F_0
 +\frac{3}{8} (1-\cos \theta_e^*)^2 F_L
 +\frac{3}{8} (1+\cos \theta_e^*)^2 F_R
\Bigg] \; ,
\eeq
where $F_0+F_L+F_R=0$.
Recent measurements of the $W$-boson helicity fraction have been presented by 
CDF~\cite{Aaltonen:2012hy}, D0~\cite{Abazov:2010jn}, CMS~\cite{CMS-PAS-TOP-11-020} and
ATLAS~\cite{Aad:2012ky} and CDF and D0 combined~\cite{Aaltonen:2012rz}.

Data on single top production also constrains the form of the $Wtb$ vertex \cite{Abazov:2011pm,:2012iwa}.
The majority of the data has so far been used to constrain the normalization 
and to place a constraint on the CKM mixing and $V_{tb}$~\cite{Aad:2012ux}. 

\section{Theory of non-standard couplings\protect\footnote{Authors: Yang Bai, University of Wisconsin and Matthew Baumgart, Carnegie Mellon}}

\subsection{General considerations}

The effects of new physics at a high scale $\Lambda \gg v\equiv 246$~GeV can be described by an effective Lagrangian,
\beq
{\cal L}^{\rm eff} = \sum \frac{C_i}{\Lambda^2}\,\mathcal{O}_i \,+\, \cdots \,,
\eeq
where the $\mathcal{O}_i$ are higher dimension operators.  The leading contributions enter at dimension six.  
They include corrections to $t \bar t V$ gauge vertices, four-fermion operators, and couplings involving gluons that will appear in $t \bar t$ production.  
The vertex function approach \cite{Boos:1999dd,Baur:2004uw,Baur:2005wi,Baur:2006ck} can be mapped into the operator method as we discuss in Section \ref{subsec:tw}.  The latter has the 
advantage of generality and can be used with off-shell particles and in loop calculations.  Since the prefactor for our dimension six operators involving the Higgs is $v^2/\Lambda^2$ or ($m_t \, v/\Lambda^2$) instead of $E^2/\Lambda^2$, many of these do 
not decouple at low energies.  This is an experimental boon as it allows us to probe new physics in the top sector with low energy experiments, as we describe in Section \ref{sec:limits}.  The top couplings often enter the relevant observables at loop 
level, and thus effective field theory is the theoretically consistent description.  
  
Ref.~\cite{AguilarSaavedra:2008zc} reduces an 
overcomplete basis of operators using equations of motion and gauge invariance.  However, only processes affecting the top quark's interaction with gauge 
bosons were considered.  Additional modifications of single and doubly produced tops from four-fermion operators are given in \cite{Zhang:2010dr}.  
Its authors mainly consider $t \bar t$ produced in the color octet state as only this will interfere with the SM. Also given in \cite{Zhang:2010dr} are 
higher dimension operators without tops that nonetheless affect $t \bar t$ production.  To this we add our own list of operators relevant for 
$tt$ (distinct from $t \bar t$) processes, which could also contribute to color singlet $t \bar t$.  
\begin{table}[htdp]
\caption{Dimension-6 operators for $t \bar t$ production, single $t$ production, $t$ decay, and $tt$ production.  The $q$ and $u$ fields without flavor superscript are third generation, except for the neutral current and $tt$ sections where $i$ and $j$ run from 1-3.  
As usual, $\tilde{\phi} = \epsilon \phi^*.$}
\begin{center}
\begin{tabular}{|c|c|}
\hline
\rowcolor[gray]{.9}
Charged current single top production and top decay & \\ \hline
$ \mathcal{O}^{(3)}_{\phi q} = i (\phi^\dagger \tau^I D_\mu \phi) (\bar{q} \gamma^{\mu} \tau^I q) $ & 
$ \mathcal{O}_{\phi \phi} = i (\tilde{\phi}^\dagger D_\mu \phi) (\bar{u} \gamma^\mu d) $ \\ \hline
$\mathcal{O}_{uW} = (\bar{q} \tau^I \sigma^{\mu\nu} u) \tilde{\phi} W^I_{\mu\nu} $ &
$\mathcal{O}_{dW} = (\bar{q} \tau^I \sigma^{\mu\nu} d) \phi W^I_{\mu\nu} $ \\ \hline
\rowcolor[gray]{.9}
Neutral current top production and top decay & \\ \hline
$\mathcal{O}^{(1)}_{\phi q} = i (\phi^\dagger D_\mu \phi) (\bar{q}^i \gamma^\mu q^j) $ & 
$\mathcal{O}_{\phi u} = i (\phi^\dagger D_\mu \phi) (\bar{u}^i \gamma^\mu u^j) $ \\ \hline
$ \mathcal{O}_{uB} = (\bar{q}^i \sigma^{\mu\nu} u^j) \tilde{\phi} B_{\mu\nu}  $   & \\ \hline
\rowcolor[gray]{.9}
Single top and $t \bar t$ production & \\ \hline
$\mathcal{O}_{uG} = (\bar{q} \lambda^a \sigma^{\mu\nu} u) \tilde{\phi} G^a_{\mu\nu} $ & 
$\mathcal{O}^{(1,3)}_{qq} =   (\bar{q}^i \gamma_{\mu} \tau^I q^j) (\bar{q} \gamma^{\mu} \tau^I q)$  \\ \hline
\rowcolor[gray]{.9}
$t \bar t$ production & \\ \hline
$\mathcal{O}^{(8,1)}_{qq} =   (\bar{q}^i \gamma_\mu \lambda^a q^j) (\bar{q} \gamma^\mu \lambda^a q)$ &
$\mathcal{O}^{(8,3)}_{qq} =   (\bar{q}^i \gamma_\mu \lambda^a \tau^I q^j) (\bar{q} \gamma^\mu \lambda^a \tau^I q)$ \\ \hline
$\mathcal{O}^{(8)}_{ut} =   (\bar{u}^i \gamma_\mu \lambda^a u^j) (\bar{u} \gamma^\mu \lambda^a u)$ &
$\mathcal{O}^{(8)}_{dt} =   (\bar{d}^i \gamma_\mu \lambda^a d^j) (\bar{u} \gamma^\mu \lambda^a u)$ \\ \hline
$\mathcal{O}^{(1)}_{quS} = (\bar{q} u^i) (\bar{u}^j q)$ & 
$\mathcal{O}^{(1)}_{qdS} = (\bar{q} d^i) (\bar{d}^j q)$ \\ \hline
$\mathcal{O}^{(1)}_{qtS} = (\bar{q} u) (\bar{u} q)$ & \\ \hline
\rowcolor[gray]{.9}
Gluon operators that affect $t \bar t$ production & \\ \hline
$\mathcal{O}_G = f_{ABC} G^{A\nu}_\mu G^{B\rho}_\nu G^{C\mu}_\rho$ &
$\mathcal{O}_{\tilde{G}} = f_{ABC} \tilde{G}^{A\nu}_\mu G^{B\rho}_\nu G^{C\mu}_\rho$ \\ \hline
$\mathcal{O}_{\phi G} = \phi^\dagger \phi \, G^A_{\mu\nu} G^{A\mu\nu} $ &
$\mathcal{O}_{\phi\tilde{G}} = \phi^\dagger \phi \, \tilde{G}^A_{\mu\nu} G^{A\mu\nu} $ \\ \hline
$\mathcal{O}_{GB} = G^{A\nu}_\mu \tilde{G}^{A\rho}_\nu B^{C\mu}_\rho$ & \\ \hline
\rowcolor[gray]{.9}
$tt$ production and color singlet $t \bar t$ production & \\ \hline
$\mathcal{O}_{qqV}^{(1)} = (\bar{q}^i \gamma_\mu q^j) (\bar{q}^k \gamma^\mu q^l)$ &
$\mathcal{O}^{(3)}_{qq} =   (\bar{q}^i \gamma_\mu \tau^I q^j) (\bar{q}^k \gamma^\mu \tau^I q^l)$ \\ \hline
$\mathcal{O}_{quV}^{(1)} = (\bar{q}^i \gamma_\mu q^j) (\bar{u}^k \gamma^\mu u^l)$ & 
$\mathcal{O}_{uuV}^{(1)} = (\bar{u}^i \gamma_\mu u^j) (\bar{u}^k \gamma^\mu u^l)$ \\ \hline
\rowcolor[gray]{.9}
$t \bar t h$ coupling & \\ \hline
${\cal O}_{3\phi}=\phi^\dagger \phi \, \tilde{\phi} \, \bar q \, u$ & \\ \hline
\end{tabular}
\end{center}
\label{tbl:dimsix}
\end{table}%

%A Minimal set of top anomalous couplings\cite{AguilarSaavedra:2008zc}

%QCD Corrections to Flavor Changing Neutral Coupling
%                        Mediated Rare Top Quark Decays\cite{Drobnak:2010by}.

%Constraints on Non-standard Top Quark Couplings\cite{Zhang:2012cd}

%Anomalous Top Couplings at Hadron Colliders Revisited\cite{Bach:2012fb}

%Optimal-observable Analysis of Possible Non-standard
%                        Top-quark Couplings in $pp \to t \bar{t} X \to l^+$\cite{Hioki:2012vn}

%Top electroweak couplings\cite{Baur:2004uw,Baur:2005wi,Baur:2006ck}

%Effective Field Theory for Nonstandard Top Quark
%                        Couplings\cite{Greiner:2011tt}

%Quark mixing: Determination of top couplings\cite{delAguila:1999ka}

\subsection{$t W$ couplings}
\label{subsec:tw}

The first four operators  can affect the $Wtb$ couplings. To match the traditional formulas in terms of form factors defined as
\beq
\mathcal{L}_{int} \supset - \frac{g}{\sqrt{2}} \bar{b}\gamma^\mu \left( c^W_L P_L + c^W_R P_R\right)\,t\,W^-_\mu 
- \frac{g}{\sqrt{2}} \,\bar{b}\, \frac{i\sigma^{\mu\nu} q_\nu}{M_W} \left( d^W_L P_L + d^W_R P_R\right) \, t\, W^-_\mu \,+\, h.c., 
\label{eqn:anom}
\eeq
where $q$ is the $W$ momentum and $c^W_L$ equals the CKM matrix element $V_{tb}\approx 1$ and $c^W_R, d^W_L, d^W_R$ vanish at the tree level in the SM.  We have the modifications of these form factors from the four dimension-six operators~\cite{AguilarSaavedra:2008zc}:
\begin{align}
&\delta c^W_L = C^{(3)*}_{\phi q}\,\frac{v^2}{\Lambda^2} \,,  
&\delta d^W_L = \sqrt{2} \, C_{dW} \frac{v^2}{\Lambda^2} \,, \\
&\delta c^W_R = \frac{1}{2}\,C^{(3)*}_{\phi \phi}\,\frac{v^2}{\Lambda^2} \,, 
&\delta d^W_R = \sqrt{2}\, C_{uW} \frac{v^2}{\Lambda^2} \,.
\end{align}
In principle, the form factors should be functions of the momentum $q$. The additional $q^2$ terms in the form factors will match to higher-dimensional operators ({\it e.g.~}dimension-8 operators).   Since there are two expansion parameters, $v^2/\Lambda^2$ and $q^2/\Lambda^2$, to use the effective operators, one needs to be cautious and should not have the momentum in the process to be above $\Lambda$ as stressed in~\cite{Baur:2004uw}. Existing collider studies on 
probing anomalous $W t b$ couplings in single top production can be found in \cite{Boos:1999dd}. 

%\subsubsection{$W$ helicity fractions}
%W polarisation beyond helicity fractions in top quark
%                        decays\cite{AguilarSaavedra:2010nx}

\subsection{$t\, Z$, $t \gamma$ and $t g$ couplings}
\label{subsec:tZa}

The $t\bar t Z$ couplings are similar to those involving the $W$.  The vertices take the form 
\beqn
\mathcal{L}_{int} \supset -\frac{g}{2\,c_W}\,\bar{t} \,\gamma^\mu \left( c^Z_L P_L + c^Z_L P_R - 2\,s_W^2\,Q_t  \right)\,t\,Z_\mu \,-\,\frac{g}{2\,c_W}\,\bar{t} \,\frac{i\sigma^{\mu\nu} q_\nu}{M_Z} \left( d^Z_V \,+\, i d^Z_A \gamma_5 \right)\,t \,Z_\mu \,,
\eeqn
with $Q_t=2/3$ as the electric charge of the top quark and $c_W=\cos{\theta_W}$. In the SM, the couplings are $c^Z_L=1$, $c^Z_R=0$ and $d^Z_{V, A}=0$ at the tree level. The match to the dimension-six operators is given by
\begin{align}
&\delta c^Z_L = \mbox{Re}\left[ C^{(3)}_{\phi q} - C^{(1)}_{\phi q} \right]\,\frac{v^2}{\Lambda^2} \,,
&\delta d^Z_V = \sqrt{2} \,\mbox{Re}\left[c_W C_{uW} - s_W C_{uB} \right] \frac{v^2}{\Lambda^2} \,, \\
&\delta c^Z_R = - \mbox{Re} \left[C_{\phi u}\right]\,
\frac{v^2}{\Lambda^2} \,,  
&\delta d^W_R = \sqrt{2}\, \mbox{Im}\left[c_W C_{uW} - s_W C_{uB} \right]  \frac{v^2}{\Lambda^2} \,.
\end{align}

For the $t \bar t \gamma$ couplings, the vertices are
\beqn
\mathcal{L}_{int} \supset -e\,Q_t\,\bar{t}\gamma^\mu t \,A_\mu \,-\, e\,\bar{t}\, \frac{i\sigma^{\mu\nu} q_\nu}{m_t}\left( d^\gamma_V + i d^\gamma_A \gamma_5\right)t\,A_\mu \,. 
\eeqn
Here, the couplings $d^\gamma_V$ and $d^\gamma_A$ are related to the top quark magnetic and electric dipole moment, respectively. The match to the dimension-six operators is given by
\begin{align}
&\delta d^\gamma_V = \frac{\sqrt{2}}{e}\, \mbox{Re}\left[ c_W\,C_{u B} + s_W\,C_{uW} \right]\,\frac{v\,m_t}{\Lambda^2} \,,
&\delta d^\gamma_A = \frac{\sqrt{2}}{e}\, \mbox{Im}\left[ c_W\,C_{u B} + s_W\,C_{uW} \right]\,\frac{v\,m_t}{\Lambda^2} \,.
\end{align}
To measure the $t \bar t \gamma$ couplings, 
it is important to know the SM  next-to-leading order QCD corrections \cite{Melnikov:2011ta}.

For the $t \bar t g$ couplings, the vertices are
\beqn
\mathcal{L}_{int} \supset -g_s\,\bar{t}\,\frac{\lambda^a}{2}\,\gamma^\mu t \,G^a_\mu \,-\, g_s\,\bar{t}\, \gamma^a\,\frac{i\sigma^{\mu\nu} q_\nu}{m_t}\left( d^g_V + i d^g_A \gamma_5\right)t\,G^a_\mu \,. 
\eeqn
Here, the couplings $d^g_V$ and $d^g_A$ are related to the top quark chromo-magnetic and chromo-electric dipole moment, respectively. The match to the dimension-six operators is given by
\begin{align}
&\delta d^g_V = \frac{\sqrt{2}}{g_s}\, \mbox{Re}\left[ C_{u G}  \right]\,\frac{v\,m_t}{\Lambda^2} \,,
&\delta d^g_A = \frac{\sqrt{2}}{g_s}\, \mbox{Im}\left[ C_{uG} \right]\,\frac{v\,m_t}{\Lambda^2} \,.
\end{align}

\subsection{$t H$ couplings}
\label{subsec:th}

The $t\bar t h$ vertex is given by
\beqn
\mathcal{L}_{int} \supset -\,\frac{\sqrt{2}\, m_t}{v}\,c^h_t\, h\,\bar{t}\,t - c^{h5}_t\,h\,\bar{t}\,i\,\gamma_5 t\,.
\eeqn
In the SM, $c^h_t=1$ and $c^{h5}_t=0$. With the new physics from dimension-six operators, we have
\begin{align}
&\delta c^h_t = -\,\frac{3}{\sqrt{2}}\,\mbox{Re}C_{3\phi} \,\frac{v^2}{\Lambda^2} \,, & 
\delta c^{h5}_t = \left( \mbox{Im} C^{(3)}_{\phi q} - \mbox{Im} C^{(1)}_{\phi q}  + \mbox{Im} C^{(1)}_{\phi u}   \right) \frac{v\,m_t}{\Lambda^2} \,.
\end{align}
Additional interactions like $t \bar t h g$ can be generated by the ${\cal O}_{u G}$ operator and will affect the $t \bar t + h$ production cross sections. Collider studies on measuring the top Yukawa coupling at the ILC at $\sqrt{s}=500$ GeV can be found in \cite{Yonamine:2011jg}. Associated production of Higgs and single top at hadron colliders can be found in \cite{Maltoni:2001hu,Barger:2009ky}.

\subsection{Top Quark Couplings to Dark Matter}
Assuming dark matter is a weak scale particle, we can make a list of effective operators for the top quark couplings to $\chi$. For a Dirac fermion $\chi$, we have~\cite{Bai:2010hh}
\beq
\mathcal{L}_{int} \supset \frac{1}{\Lambda_1^2} \bar{\chi} \chi \,\bar{t} t \, +\, \frac{1}{\Lambda_2^2} \bar{\chi} \gamma_5\chi \,\bar{t} \gamma^5 t \,+\, \frac{1}{\Lambda_3^2} \bar{\chi}\gamma^\mu \chi \,\bar{t} \gamma_\mu t \,+\,\frac{1}{\Lambda_4^2} \bar{\chi} \gamma_\mu \gamma_5\chi \,\bar{t} \gamma^\mu \gamma_5 t \,+\, h.c.\,.   
\eeq
The signal could be $\bar{t}t+\mbox{MET}$. One can also close the top quark loop and discuss mono-jet + MET at a hadron collider~\cite{Haisch:2012kf} and/or mono-photon+ MET at a linear collider.

\subsection{Constraints on non-standard couplings}
\label{sec:limits}

Several papers have placed constraints on either the coefficients of the operators in Table \ref{tbl:dimsix} and/or the couplings in Sections~\ref{subsec:tw}-\ref{subsec:th}.  In general, the limits are placed on couplings that affect 
$t \bar t V$ and $Wtb$ vertices.  In some cases, the limits involve actual data, in others they are hypothetical, invoking either future LHC data or the ILC.  In \cite{Greiner:2011tt} and \cite{Zhang:2012cd}, 
the authors follow up on the basis written down in \cite{Zhang:2010dr}.  Along with the helicity fraction in top decays, they use precision electroweak data and results from $\bar B \rightarrow X_s \gamma$ and $B - \bar B$ mixing 
to constrain the nine dimension-6 operators that couple 3$^{\rm rd}$ generation quarks to electroweak bosons, a list that strongly overlaps with Table \ref{tbl:dimsix}.  $B$-physics observables are also used to place limits on $Wtb$ couplings in \cite{Drobnak:2011aa}.
Refs.~\cite{Bach:2012fb, Baur:2004uw} discuss possible bounds on $Wtb$ at the LHC.  In \cite{Baur:2005wi,Baur:2006ck}, future LHC data is simulated along with the ILC for constraining top couplings to vector bosons.  In \cite{AguilarSaavedra:2012vh}, the operator basis is discussed in 
the context of the ILC along with the claim that beam polarization and a CM energy of 1 TeV would allow one to separately probe all the operators contributing to $t \bar t Z$ and $t \bar t \gamma$ couplings.  Lastly, by using alternative $W$ helicity variables, ref.~\cite{AguilarSaavedra:2010nx} proposes to measure phases in the $Wtb$ couplings that are otherwise difficult to access.

In addition to discussing electroweak couplings, \cite{Hioki:2012vn} uses Tevatron and LHC data on the overall $t \bar t$ cross section to constrain the top's chromomagnetic and electric dipole moments (respectively given by the real and imaginary coefficients of $\mathcal{O}_{uG}$).  In addition to collider data, \cite{Kamenik:2011dk} uses neutron and mercury EDMs along with $B$ decays to place limits on the top's chromo and regular dipole moments.  Ref.~\cite{Baumgart:2012ay} uses top quark spin correlations in a complementary approach to constrain the top's color dipole moments. 
For those operators involving neutral currents, ref.~\cite{Drobnak:2010by} uses CDF and ZEUS data to place constraints on $t \rightarrow Z q$ and $t \rightarrow \gamma \, q$ decays.  Also, \cite{delAguila:1999ka} discusses the potential reach of the LHC to study FCNC top decays. As more LHC data becomes available, one can constrain more of the operators in Table \ref{tbl:dimsix}.  The overall $t \bar t$ rate will place limits on the ``$t \bar t$ production'' set, including many four quark operators that have yet to be significantly bounded.  Additionally, $tt$ or 4 top ({\it cf.}~\cite{Lillie:2007hd}) events could give additional information on the four fermion subset of Table \ref{tbl:dimsix}.  Lastly, the direct observation of $t \bar t h$ events will allow us to better determine the coefficients of the many operators that couple tops and Higgses.

%% file: Alvarez/Alvarez.tex
\begin{figure}[!h!tpb]
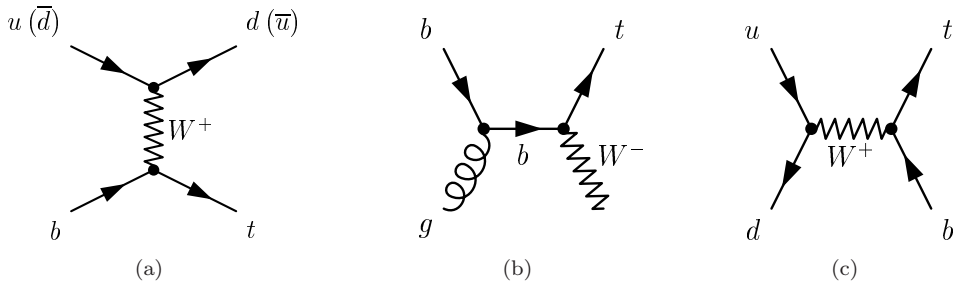

  \centering
  \subfigure[]{
    \includegraphics[width=0.25\textwidth]{Alvarez/singTopWg3Pict.epsi}
    \label{fig:tchan}
  }
  \hspace*{0.07\textwidth}
  \subfigure[]{
    \includegraphics[width=0.18\textwidth]{Alvarez/singTopAssocPict.epsi}
    \label{fig:Wt}
  }
  \hspace*{0.07\textwidth}
  \subfigure[]{
    \includegraphics[width=0.18\textwidth]{Alvarez/singTopSChanPict.epsi}
    \label{fig:schan}
  } 
\caption{Feynman diagrams of single top-quark production processes.}
%         \subref{fig:tchan} $t$-channel production, 
%         \subref{fig:Wt} associated $Wt$ production, and
%         \subref{fig:schan} $s$-channel production.}
\label{fig:feynman}
\end{figure}

This section presents a summary of the current studies on the single top-quark production at 
the LHC and the Tevatron. 
There are multiple processes that can lead to the production of a top-quark. 
The dominant mechanism for creating top-quarks at hadron colliders is the production
of a top-antitop pair through QCD interactions.
The top-quark can also be produced singly by an electroweak $Wtb$-vertex with a smaller cross-section than 
for top-antitop production.
The SM single top-quark production proceeds through three different mechanisms shown in Fig.~\ref{fig:feynman}: 
$t$-channel  exchange of a $W$-boson, associated production of a top-quark and a $W$ boson ($Wt$-channel), 
and $s$-channel production and decay of a virtual $W$~boson. The theoretical cross-section predictions 
for these processes at different center-of-mass energies are given in Table~\ref{tab_xsection}.

 \begin{table}[h]\vspace{-.2cm}
   \begin{center}\small
     \begin{tabular}{lccc}
       \hline \hline
       $E_{CM}$ [\tev]    & $t$-channel & $Wt$-channel  & $s$-channel   \\ 
       \hline \hline
       1.96   & 2.10$\pm$0.19 pb         & 0.22$\pm$0.08 pb          & 1.05$\pm$0.07 pb\\
       7      & 64.57$^{+3.32}_{-2.62}$ pb & 15.74$^{+1.34}_{-1.36}$ pb  & 4.63$^{+0.29} _{-0.27}$ pb \\
       8      & 87.8$\pm$3.4 pb          & 22.4$\pm$1.5 pb           & 5.6$\pm$0.3 pb\\
       \hline \hline     
       \end{tabular}       
     \end{center} \vspace{-.3cm} 
  % \caption{NNNLO and NNLO single top-quark production cross-sections at different center-of-mass energies~\cite{Kidonakis:tevatron,Kidonakis:tchannel,Kidonakis:wt,Kidonakis:schannel} with $m_{top}$=172.5~GeV/c$^2$.}  
\caption{Approximate next-to-next-to-next-to-leading (NNNLO) and next-to-next-to-leading (NNLO) single top-quark production cross-sections at different center-of-mass energies~\cite{Kidonakis:tevatron,Kidonakis:tchannel,Kidonakis:wt,Kidonakis:schannel} with m$_{top}$=172.5 GeV/c$^2$.
The Tevatron (1.96~\tev) cross sections are calculated at approximate NNNLO.  The LHC (7 and 8~\tev) cross sections are calculated at approximate NNLO.}
   \label{tab_xsection}
 \end{table}
There are several motivations to study single top production: 
single top production gives complementary information 
on top-quark properties, it allows a direct measurement of the CKM matrix element $V_{tb}$,
and is sensitive to many models of new physics. Furthermore, determining the cross-section also allows to extract the $b$-quark
density.

This section is organized as follows: $t$-channel and $Wt$-channel production cross-section measurements including
the determination of $V_{tb}$ are presented, followed by a review of searches in the $tb$ final states 
such as $s$-channel and $W'$.

\subsection{$t$-channel single top production}
The $t$-channel single top production is dominant both at the Tevatron and at the LHC.
The measurements of the $t$-channel single top-quark production cross-section with the ATLAS detector 
are done using a neural network based discriminant. 
Selected events in the lepton+jets channel contain one lepton, missing transverse momentum, and two or three jets, 
including one which is $b$-tagged. 
The $t$-channel production cross-section is measured by performing a combined binned maximum likelihood fit to the neural 
network output distribution for the observed data resulting in $\sigma_t$=95$\pm$18~pb 
with 5.8 fb$^{-1}$ at 8~\tev~\cite{ATLAStchan8TeV} and  $\sigma_t$=83$\pm$4(stat) $^{+20}_{-19}$(syst)~pb 
using 1.04 fb$^{-1}$ at 7~\tev~\cite{ATLAStchan7TeV}, which are in good agreement with the SM prediction. 
Major systematic uncertainties considered include jet energy scale (JES), $b$-tagging efficiency, and the amount of initial- and final-state 
radiation. Using the ratio of the measured to the predicted cross-section and assuming that the 
top-quark-related CKM matrix elements obey the relation $|V_{tb}|>> |V_{ts}|$, $|V_{td}|$, the coupling strength at the $W-t-b$ 
vertex is determined to be $|V_{tb}|$=1.04$^{+0.10}_{-0.11}$ at 8~\tev~and $|V_{tb}|$ = 1.13$^{+0.14}_{-0.13}$ at 7~\tev. 
If it is assumed that $|V_{tb}|\leq$1, a lower limit of $|V_{tb}|>$0.80 (0.75) at 8 (7)~\tev~is obtained at 95$\%$ CL.
With the same selection described above and the same analysis technique the top-quark and top-antiquark production cross-sections 
are also measured using an integrated luminosity of 4.7 fb$^{-1}$ at 7~\tev. A binned maximum likelihood fit to the output 
distribution of neural networks 
that are split according to the charge of the lepton is performed resulting in $\sigma_t(t)$=53.2$\pm$10.8~pb and 
$\sigma_t(\bar{t})$=29.5$^{+7.4}_{-7.5}$~pb~\cite{ATLASratiotchan7TeV}. A cross-section ratio of R$_t$=1.81$^{+0.23}_{-0.22}$ is measured. 
At the current precision, the measurement 
is in agreement with the predictions based on various global PDF sets that range from 1.86 to 2.07.

The CMS strategies for the $t$-channel single top cross-section analyses at 7 and 8~\tev~are different.
The 8~\tev~result uses only the leptonic decay channel with a muon in the final state. 
%After a selection optimized for the $t$-channel production, 
The pseudorapidity distribution of the recoil jet is exploited for the measurement. 
%The main backgrounds are estimated with data driven techniques. 
With an integrated luminosity of 5.0 fb$^{-1}$ the cross-section is found to 
be 80.1$\pm$5.7(stat)$\pm$11.0(syst)$\pm$4.0(lumi) ~pb~\cite{CMStchan8TeV}.
The largest systematic uncertainties are from statistical uncertainty, JES and $t$-channel generator uncertainties.
$|V_{tb}|$ is also measured at the 10$\%$ level obtaining a lower bound of $|V_{tb}| >$ 0.81 at 95$\%$ CL.
The CMS results at 7 \tev~followed two different and complementary approaches. The first approach 
exploits the distributions of the pseudorapidity of the recoil jet and reconstructed top-quark mass.
%using background estimates determined from control samples in data. 
The second approach is based on multivariate analysis techniques. 
%that probe the compatibility of the candidate events with the signal. 
In this case, both the muon and the electron decay channels have been used, 
corresponding to integrated luminosities of 1.17 and 1.56 fb$^{-1}$, respectively. The single top-quark production 
cross-section in the $t$-channel is measured to be 67.2$\pm$6.1~pb~\cite{CMStchan7TeV}.
The largest systematic uncertainties are from statistical uncertainty, W+jets normalization and generator uncertainties.
A lower limit of $|V_{tb}|>$ 0.92 is obtained at 95$\%$ CL using the SM assumption of $|V_{tb}|<$ 1.

Following the first observation of 
single top-quark production in 2009 at the Tevatron~\cite{CDFobssingletop, D0obssingletop, Fermiobssingletop}, 
both collaborations, CDF and D0, have updated their single top results.
D0 presents measurements of production cross sections of single top-quarks in $p\bar{p}$ collisions at 
1.96~\tev~in a data sample corresponding to an integrated luminosity of 5.4 fb$^{-1}$.
Selected events with an isolated electron or muon, an imbalance in transverse energy, and two, three, 
or four jets, with one or two of them containing a $b$-quark. Three different multivariate analysis techniques 
to extract the signal are used in this analysis. The measured inclusive cross section, for $t$-channel 
and $s$-channel together, 
is $\sigma$=3.43$_{-0.74}^{+0.73}$ pb and it used to extract the CKM matrix element 0.79$<|V_{tb}|\leq$1 
at the 95$\%$ CL. The largest uncertainties arise from the JES, JER, 
corrections to $b$-tagging efficiencies, and the correction for jet-flavor composition in W+jets events.
The cross sections are also measured separately to be $\sigma_s$=0.68$_{-0.35}^{+0.38}$ pb 
and $\sigma_t$=2.86$_{-0.63}^{+0.69}$ pb~\cite{D0singletop}, assuming, respectively, $t$-channel and $s$-channel 
production rates as predicted by the SM.

A measurement of single top-quark production in lepton+jets final state using 7.5 fb$^{-1}$
of data collected by CDF Run II experiment is also presented. Selected events 
contain a charged lepton, electron or muon, missing transverse energy and two or three jets, 
at least one of them $b$-tagged. 
A Neural Network multivariate method is used to discriminate signal against comparatively large backgrounds. 
The measured inclusive cross-section is 3.04$^{+0.57}_{-0.53}$~pb (stat+syst)~\cite{CDFsingletop} 
and the CKM matrix element value $V_{tb}$=0.96$\pm$0.09(stat+syst)$\pm$0.05(theory) with a lower limit of 
$|V_{tb}| >$ 0.78 at the 95$\%$ CL. The systematic uncertainty source with the biggest contribution is the background normalization.
The $s$-channel and $t$-channel cross sections are also extracted separately performing a two dimensional fit, 
$\sigma_s$= 1.81$^{+0.63}_{-0.58}$ pb and  $\sigma_t$ = 1.49$^{+0.47}_{-0.42}$ pb.

\subsection{$Wt$-channel single top production}
The $Wt$-channel is negligible at the Tevatron but it has the second highest cross-section among single top processes at the LHC.
The CMS and ATLAS collaborations have shown evidence for the associated production of a W boson and a top-quark.
The CMS measurement is performed using events with two leptons and a jet originating from a $b$-quark. A multivariate analysis 
based on kinematic properties is exploited to separate the main background contribution, $t\bar{t}$, from the signal. The observed signal has a 
significance of 4.0$\sigma$ and corresponds to a cross-section of 16$^{+5}_{-4}$~pb~\cite{CMSWt7TeV} with  4.9 fb$^{-1}$at  7 \tev.
The measurement can be used to determine the absolute value of the CKM matrix element $|V_{tb}|$ = 1.01$^{+0.16}_{-0.13}$(exp)$^{+0.03}_{-0.04}$(theo) 
assuming $|V_{ts}|$ and $|V_{td}|$ much smaller than $|V_{tb}|$.

The ATLAS $Wt$-channel analysis is based on the selection of the dileptonic final states with events featuring two isolated
leptons, electron or muon, with significant transverse missing momentum and at least one jet. 
A template fit to a boosted decision tree output distribution is performed to determine the cross-section. 
The result is incompatible with the background-only hypothesis at the 3.3$\sigma$ level.
The corresponding cross-section is determined and found to be $\sigma_{Wt}$=16.8$\pm$2.9(stat)$\pm$4.9(syst) ~pb~\cite{ATLASWt7TeV} using
2.05 fb$^{-1}$ at 7 \tev. The CKM matrix element $|V_{tb}|$ = 1.03$^{+0.16}_{-0.19}$ is derived assuming that the $Wt$-channel 
production through $|V_{ts}|$ and $|V_{td}|$ is small.
The results based on the lepton+jets final state have been investigated however their contributions will result in a smaller sensitivity.

\subsection{$s$-channel single top production}
Searches in the $tb$ final state are particularly interesting since the SM $s$-channel production mode 
itself has not been observed yet and the search for this process is sensitive to several models of new 
physics~\cite{schanTait:2000sh}. 
A search for $s$-channel top-quark production has been performed using  
0.7~fb$^{-1}$ of ATLAS data at a center-of-mass energy of 7 TeV. Selected events contain one lepton, missing transverse energy and two jets. 
The final selection requires both jets to be identified as coming from $b$-quarks. 
An observed upper limit at 95$\%$ CL on the $s$-channel single top-quark production cross-section of $\sigma_{s}<$ 26.5 ~pb~\cite{ATLASschannel7TeV}, which corresponds to about 5 times the signal SM cross-section,
is obtained using a cut-based analysis.

One can also search for new heavy gauge bosons such as the $W'$ boson by looking for $tb$ resonances.
The most stringent limits on a right-handed $W'_R$ with SM-like couplings in the decay mode $W'_R \rightarrow tb$ 
are set by the CMS experiment and excludes a $W'_R$ boson mass below 1.85~\tev~at 95$\%$ CL~\cite{CMSWprime7TeV} using 
5 fb$^{-1}$ of data at 7 \tev.
The  $W'_R \rightarrow tb$  decay channel has been also searched for at ATLAS~\cite{ATLASWprime7TeV} and at the Tevatron~\cite{Wprime_D0,Wprime_CDF}. 

The precision achieved in the cross-section measurements 
at 7 and 8 \tev~is comparable with theoretical NLO and NNLO predictions.
A combination of the results will provide a more precise measurements 
and also stronger constraints on the CKM matrix element $|V_{tb}|$.
The evolution of the single top cross sections with energy are shown in 
Fig.~\ref{xsec}.

%% file: Loginov/symbols.tex
%run I gg + met search
\newcommand{\sla}[1]{/\!\!\!#1}
\newcommand{\mettsm}{\mbox{\scriptsize ${\rm \not\! E}_{\rm T}$}}
\newcommand{\mettgmh}{\mbox{${\rm \not\! E}_{\rm T}>35$~GeV}}
\def\absvalue#1{{|#1|}}
\newcommand{\mean}[1]{\langle #1 \rangle}
\def \mc     {\multicolumn}
\let\absv=\absvalue
\def \ie {\it i.e.}
\def\absq#1{{{|#1|}^2}}
\def \Et {{\rm E}_{\rm T}}
\def \Kt {{{\rm K}_{\rm T}}}
\def \kt {{{\rm k}_{\rm T}}}
\newcommand{\ET}{E$_{\rm T}$}
\def \Ete {{\rm E}_{\rm T}^e}
\def \Eta {{\rm E}_{\rm T}^{\gamma}}
\def \Etmu {{\rm E}_{\rm T}^{\mu}}
\def \Etnu {{\rm E}_{\rm T}^{\nu}}
\newcommand{\Pt}{{\rm p}_{\rm T}}
\newcommand{\pt}{{\rm p}_{\rm T}}
\def \Pt {{\rm P}_{\rm T}}
\newcommand{\pT}{p$_{\rm T}$}
\newcommand{\met}{\mbox{${\rm \not\! E}_{\rm T}$}}
\newcommand{\Met}{\mbox{${\rm \not\! E}_{\rm T}$}} 
\newcommand{\metvec}{{\not\!\! \vec{E}_T}}
\newcommand{\lepvec}{{\vec{E}_T^\ell}}
\newcommand{\phovec}{{\vec{E}_T^\gamma}}
\def \Mt {{\rm M}_{\rm T}}
\def \Ht {{\rm H}_{\rm T}}
\def \Mtop {{\rm M}_{\rm Top}}
\def \Mll {{\rm M}_{ll}}
\def \MW {{\rm M}_{\rm W}}
\def \enu {\epsilon_{\nu}}
\def \Egamma {E$_\gamma$\,}
\newcommand{\Wenu}{\mbox{W$^\pm \rightarrow e^{\pm}\nu$}}
\newcommand{\wenu}{\mbox{W$^\pm \rightarrow e^{\pm}\nu$}}
\newcommand{\emetg}{e \met $\gamma$}
\newcommand{\mumetg}{$\mu$ \met $\gamma$}

\def \dpt{\Delta\Pt}
\def \dphi{\Delta\phi}
\def \dx{\Delta X}
\def \dy{\Delta Y}
\def \dz{\Delta Z}
\def \dr{\Delta R}
\def \dR{\Delta R}
\def \deta{\Delta \eta}

\def\lsp{\tilde{LSP}}
\newcommand{\neutralino}{\tilde\chi_{1}^{0}}
\newcommand{\chargino}{\tilde\chi_{1}^{\pm}}
\def\gluino{\tilde{g}}
\def\Goldstino{\tilde{G}}
\def\Gravitino{\tilde{G}}
\newcommand{\gravitino}{\tilde{G}}
\def\sbottom{\tilde{b}}
\def\stop{\tilde{t}}
\def\stopbar{\tilde{{\bar t}}}
\def\sbottom{\tilde{b}}
\def\sbottombar{\tilde{{\bar b}}}
\def\scharm{\tilde{c}}
\def\selectron{\tilde{e}}
\def\selectronp{\tilde{e}^+}
\def\selectronm{\tilde{e}^-}
\def\sneutrino{\tilde{\nu}}
\def\smuon{\tilde{\mu}}
\def\squark{\tilde{q}}
\def\stau{\tilde{\tau}}
\def\stauplus{\tilde{\tau}^+}
\def\stauminus{\tilde{\tau}^-}
\def\nustau{\tilde{\nu}_{\tau}}
\def\snu{\tilde{\nu}}
\def\snutau{\tilde{\nu}_{\tau}}
\def\snue{\tilde{\nu}_{e}}
\def\snumu{\tilde{\nu}_{\mu}}
\def\slepton{\tilde{\ell}}
\def\Soup{ {\tilde{Soup}}{\bar {\tilde{Soup}}} }
\def\Wino{\tilde{W}}
\def\wino{\tilde{W}}
\def\Zino{\tilde{Z}}
\def\zino{\tilde{Z}}
\def\Wstar{W^*}
\newcommand{\Zgstar}{Z^0\kern -0.25em/\kern -0.15em\gamma^*}
\def\Zgamma{\Z\kern -0.1em/\kern -0.1em\gamma}
\def\higgszero{H^0}
\def\higgsplus{H^+}
\def\photino{\tilde{\gamma}}
\def\pizero{\pi ^0}
\def\eeggmet{ee\gamma\gamma\Met}
\def\llgg{\ell\ell\gamma\gamma}
\def\gmet{\gamma\Met}
\def\llggmet{\ell\ell\gamma\gamma\Met}
\newcommand{\lgX}{\ell\gamma+X}
\newcommand{\egX}{e\gamma+X}
\newcommand{\mugX}{\mu\gamma+X}
\newcommand{\ljX}{\ell j+X}
\newcommand{\ejX}{e j+X}
\newcommand{\mujX}{\mu j+X}
\newcommand{\leX}{\ell e+X}
\newcommand{\eeX}{e e+X}
\newcommand{\mueX}{\mu e+X}
\def\lgmet{\ell\gamma\Met}
\def\lgmetb{\ell\gamma\Met b}
\def\lgbmet{\ell\gamma\Met b}
\def\ljmet{\ell j\Met}
\def\ljmetb{\ell j\Met b}
\def\lemetb{\ell e\Met b}
\def\lgal{\ell\gamma}
\def\llg{\ell\ell\gamma}
\def\llj{\ell\ell j}
\def\lgg{\ell\gamma\gamma}
\def\ggX{\gamma\gamma+X}
\def\ggmet{\gamma\gamma\met}
\def\ggg{\gamma\gamma\gamma}
\def\gggg{\gamma\gamma\gamma\gamma}
\def\ll{\ell\ell}
\def\egmet{e\gamma\Met}
\def\egmetb{e\gamma\Met b}
\def\egbmet{e\gamma\Met b}
\def\eemet{ee\Met}
\def\eemetb{ee\Met b}
\def\ejmet{ej\Met}
\def\ejmetb{ej\Met b}
\def\mugmet{\mu\gamma\Met}
\def\mugmetb{\mu\gamma\Met b}
\def\mugbmet{\mu\gamma\Met b}
\def\mujmet{\mu j\Met}
\def\mujmetb{\mu j\Met b}
\def\muemetb{\mu e\Met b}
\def\eggmet{e\gamma\gamma\met}
\def\eegmet{ee\gamma\met}
\def\eeggmet{ee\gamma\gamma\met}
\def\mmggmet{\mu\mu\gamma\gamma\Met}
\def\mumuggmet{\mu\mu\gamma\gamma\Met}
\def\muggmet{\mu\gamma\gamma\Met}
\def\mumugmet{\mu\mu\gamma\Met}
\def\mmggjj{\mu\mu\gamma\gamma jj}
\def\egmetfj{e\gamma\Met j_{fat}}
\def\eg{e\gamma}
\def\eeg{ee\gamma}
\def\jjg{jj\gamma}
\def\jjgg{jj\gamma\gamma}
\def\bbg{bb\gamma}
\def\bbgg{bb\gamma\gamma}
\def\egg{e\gamma\gamma}
\def\eegg{ee\gamma\gamma}
\def\mumugg{\mu\mu\gamma\gamma}
\def\mg{\mu\gamma}
\def\mug{\mu\gamma}
\def\mgmet{\mu\gamma\Met}
\def\mugmet{\mu\gamma\Met}
\def\mmg{\mu\mu\gamma}
\def\mumug{\mu\mu\gamma}
\def\mumuj{\mu\mu j}
\def\mgg{\mu\gamma\gamma}
\def\mugg{\mu\gamma\gamma}
\def\mmgg{\mu\mu\gamma\gamma}
\def\lg{\ell\gamma}
\def\lmet{\ell\met}
\def\WWgg{WW\gamma\gamma}
\def\Wgg{W\gamma\gamma}
\def\Wbbar{W\bbbar}
\def\Wccbar{W\ccbar}
\def\Wc{Wc}
\def\Zgg{Z\gamma\gamma}
\def\Wggg{W\gamma\gamma\gamma}
\def\Zggg{Z\gamma\gamma\gamma}
\def\gg{\gamma\gamma}
\def\Wg{W\gamma}
\def\Zg{Z\gamma}
\def\Zeeg{Z(ee)\gamma}
\def\Zmumug{Z(\mu\mu)\gamma}
\def\Ztautaug{Z(\tau\tau)\gamma}
\def\Zeegg{Z(ee)\gamma\gamma}
\def\Zmumugg{Z(\mu\mu)\gamma\gamma}
\def\Ztautaugg{Z(\tau\tau)\gamma\gamma}
\def\Wenug{W(e\nu)\gamma}
\def\Wmunug{W(\mu\nu)\gamma}
\def\Wtaunug{W(\tau\nu)\gamma}
\def\Wenugg{W(e\nu)\gamma\gamma}
\def\Wmunugg{W(\mu\nu)\gamma\gamma}
\def\Wtaunugg{W(\tau\nu)\gamma\gamma}
\def\wg{W\gamma}
\def\zg{Z\gamma}
\def\Wj{Wj}
\def\Zj{Zj}
\def\wj{Wj}
\def\zj{Zj}
\def \intlum {\int {\cal L} dt}
\def\deg{^\circ}
\def\degrees{^\circ}
\def\gt{>}
\def\lt{<}
\def\GammaW{\Gamma(W)}
\def\GeV{GeV}
\def\GVc{GeV/c}
\def\GeVc2{GeV/{c^2}}
\def\MeV{MeV}
\newcommand{\invnb}{nb^{-1}}
\newcommand{\invpb}{pb^{-1}}
\newcommand{\bfinvpb}{\bf pb^{-1}} % Should be roman in PRL
\newcommand{\bfinvfb}{\bf fb^{-1}} % Should be roman in PRL
\newcommand{\pbinv}{pb^{-1}}
\newcommand{\invfb}{\rm fb^{-1}}
\def\pbarp{{\bar p}p}
\newcommand{\pbar}{\rm{\bar p}} % Should be roman in PRL
\newcommand{\ppbar}{p{\bar p}} % Should be roman in PRL
\def\ubar{\bar u}
\def\dbar{\bar d}
\def\cbar{\bar c}
\def\sbar{\bar s}
\def\tbar{\bar t}
\def\bbar{\bar b}
\def\uubar{u{\bar u}}
\def\ddbar{d{\bar d}}
\def\ccbar{c{\bar c}}
\def\bbbar{b{\bar b}}
\def\udbar{u{\bar d}}
\def\csbar{c{\bar s}}
\def\qbar{{\bar q}}
\def\llbar{\ell{\bar \ell}}
\def\qqbar{q{\bar q}}
\def\nunubar{\nu{\bar \nu}}
\def\ttbar{t\bar{t}}
\def\ttbarg{t\bar{t}\gamma}
\def\ttg{t\bar{t}\gamma}
\def\ttj{t\bar{t} j}
\def\ee{ee\/}
\def\mumu{\mu\mu\/}
\def\emu{e\mu\/}
\def\W{W}
\def\WW{{\em W\/\em W\ }}
\def\Wgamma{{\em W\/\em \gamma\ }}
\def\gammaWenu{{\gamma+W\rightarrow e\nu}}
\def\gammaWmunu{{\gamma+W\rightarrow \mu\nu}}
\def\gammaZee{{\gamma+Z\rightarrow ee}}
\def\gammaZmumu{{\gamma+Z\rightarrow \mu\mu}}
\def\Zeg{e+`\gamma`}
\def\zeg{e+`\gamma`}
\def\Zmug{\mu+`\gamma`}
\def\zmug{\mu+`\gamma`}
\def\WZ{{\em W\/\em Z\ }}
\def\Vtb{V_{tb}}
\def\tprime{t^{\prime}}
\def\bprime{b^{\prime}}
\def\qprime{q^{\prime}}
\def\Wprime{W^{\prime}}
\def\Zprime{Z^{\prime}}
\def\bprime{b^{\prime}}
\def\MWprime{M_{W^{\prime}}}
\def\nutau{\nu _{\tau}}
\def\Vtprimeb{V_{\tprime b}}
\def\Vtprimeq{V_{\tprime q}}
\newcommand{\goes}{\kern -0.18em\rightarrow\kern -0.18em}
\newcommand{\plus}{\kern -0.18em +\kern -0.18em}
\def\Z{Z^0}
\def \ztau   {$Z\rightarrow\tau\tau \:$}
\def\lum{{\cal L}}
\def\lumunits{{cm^{-2}s^{-1}}}
\def\epem{{\rm e^{+}e^{-}}}
\def\lplm{\ell^+\ell^-}
\def\tptm{{\tau^{+}\tau^{-}}}
\def\roots{{\sqrt s}}
\def\shat{{\hat s}}
\def\95cl{95 \%~C.L.}
\def\95CL{95 \%~C.L.}
\def\r#1 {$^{#1}$}
\def\sigW {$\sigma$(\W$\rightarrow~$e $\nu$)}
\def\sigZ {$\sigma$(\Z0$\rightarrow~\epem$)}
\def \stw {$\sin^{2}\theta_{W}$}
\hyphenation{brem-sstrah-lung proc-ess}
\def\epem{e^+e^-}
\def\mpmm{\mu^+\mu^-}
\def\epmemp{e^{+-}e^{-+}}
\def\epmepm{e^{+-}e^{+-}}
\def\mpmmpm{\mu^{+-}\mu^{+-}}
\newcommand{\scinot}[2]{{#1}$\times$10$^{#2}$}
\newcommand{\pmasym}[2]{^{+#1}_{-#2}}
%\newcommand{\pmasym}[2]{^{\mbox{\small +#1}}_{\hskip 1pt\mbox{\small --#2}}}
%
% Useful command definitions for use in thesis/other documents.
%
%---------- Distributions and scalar quantities

%---------- Particles
\newcommand{\Kz}{\mbox{$\rm K^{0}$}}
\newcommand{\pip}{\mbox{$\pi^+$}}
\newcommand{\pim}{\mbox{$\pi^-$}}
\newcommand{\pipm}{\mbox{$\pi^{\pm}$}}
\newcommand{\piz}{\mbox{$\pi^0$}}
%\newcommand{\jpsi}{\mbox{$J/\psi$\,}}
%---------- Decays
\newcommand{\wenuch}{\mbox{W$^\pm \rightarrow e^{\pm}\nu$}}
\newcommand{\Wenuch}{\mbox{W$^\pm \rightarrow e^{\pm}\nu$}}
\newcommand{\Wtaunu}{\mbox{W$^\pm \rightarrow \tau^{\pm}\nu$}}
\newcommand{\wtaunu}{\mbox{W$^\pm \rightarrow \tau^{\pm}\nu$}}
\newcommand{\Wmunu}{\mbox{W$^\pm \rightarrow \mu^{\pm}\nu$}}
\newcommand{\wmunu}{\mbox{W$^\pm \rightarrow \mu^{\pm}\nu$}}
\newcommand{\Wjj}{\mbox{W$^\pm \rightarrow jj $}}
\newcommand{\Wlnu}{\mbox{$W^\pm \rightarrow l^{\pm}\nu$}}
\newcommand{\lnu}{\mbox{$\ell\nu$}}
\newcommand{\Wtbbar}{\mbox{W$ \rightarrow t{\bar b}$}}
\newcommand{\Zbbar}{\mbox{Z$ \rightarrow b{\bar b}$}}
\newcommand{\zbbar}{\mbox{W$ \rightarrow b{\bar b}$}}
\newcommand{\Znunu}{\mbox{Z$^0 \rightarrow \nu \nu $}}
\newcommand{\Zmumu}{\mbox{Z$^0 \rightarrow \mu ^+ \mu ^-$}}
\newcommand{\zmumu}{\mbox{Z$^0 \rightarrow \mu ^+ \mu ^-$}}
\newcommand{\Ztautau}{\mbox{Z$^0 \rightarrow \tau^+\tau^-$}}
\newcommand{\ztautau}{\mbox{Z$^0 \rightarrow \tau^+\tau^-$}}
\newcommand{\Zee}{\mbox{Z$^0 \rightarrow e^+e^-$}}
\newcommand{\zee}{\mbox{$Z^0 \rightarrow e^+e^-$}}
\newcommand{\Zll}{\mbox{$Z^0 \rightarrow \ell^+\ell^-$}}
\newcommand{\zll}{\mbox{$Z^0 \rightarrow \ell^+\ell^-$}}
\newcommand{\Zjj}{\mbox{Z$^0 \rightarrow jj$}}
\newcommand{\Zqqbar}{\mbox{Z$^0 \rightarrow q{\bar q}$}}
%Supersymmetry
\newcommand{ \NI         }{ {\tilde N}_1 }
\newcommand{ \NII        }{ {\tilde N}_2 }

%---------- Misc
\newcommand{\gsim}{\mbox{\small$\stackrel{>}{\sim}$\normalsize}}
\newcommand{\lsim}{\mbox{\small$\stackrel{<}{\sim}$\normalsize}}
\newcommand{\degs}{\mbox{$^{\circ}$}}
\newcommand{\etal}{{\em et al.}}
\newcommand{\tableskip}{\vskip 5pt plus3pt minus1pt \relax}
\newcommand{\tindent}{\hskip 17pt}
\newcommand{\hfull}{\hspace*{\fill}}
\newcommand{\tline}{\protect\linebreak[4]\hfull}
\newcommand{\linespace}[1]{\protect\renewcommand{\baselinestretch}{#1}
  \footnotesize\normalsize}
%
%\def\gtsima{$\; \buildrel > \over \sim \;$}
%\def\ltsima{$\; \buildrel < \over \sim \;$}
%\def\prosima{$\; \buildrel \propto \over \sim \;$}
%\def\gsim{\lower.5ex\hbox{\gtsima}}
%\def\lsim{\lower.5ex\hbox{\ltsima}}
%\def\simgt{\lower.5ex\hbox{\gtsima}}
%\def\simlt{\lower.5ex\hbox{\ltsima}}
%\def\simpr{\lower.5ex\hbox{\prosima}}
%\def\la{\lsim}
%\def\ga{\gsim}
%
% Some generic length commands
%
\newlength{\lena}
\newlength{\lenb}
\newlength{\lenc}
\newlength{\lend}
\newlength{\tlena}
\newlength{\tlenb}
\newlength{\tlenc}
\newlength{\tlend}
\newlength{\tlene}
%
% Color commands
%
%\newcommand{\red}{\color{red}}
\newcommand{\xtiny}{\tiny}
%questions and answers:
\newcounter{myquestion}
\newcommand{\noteq}{{\vskip0.1in\bf \large \color{red} [Q: {\color{black}\addtocounter{myquestion}{1}\arabic{myquestion} }]}}
\newcommand{\noteQ}{{\bf \LARGE \color{blue} [Q: {\color{black}\addtocounter{myquestion}{1}\arabic{myquestion} }]}}
\newcommand{\notea}{{\bf \large \color{red} [A]}}
\newcommand{\noteA}{{\bf \LARGE \color{blue} [A]}}
\newcommand{\noteqq}[1]{{\bf \large \color{red} [Q: {\color{black}#1}]}}
\newcommand{\noteaa}[1]{{\bf \large \color{red} [A: {\color{black}#1}]}}
%\newcommand{\noteq}{{\bf \large \color{red} [Q]}}
%from Toback's paper on GMSB diphotons+met
\newcommand{\ett}{\Et}
\newcommand{\ptt}{\mbox{$p_T$}}
\newcommand{\NONE}{\mbox{$\widetilde{\chi}_1^0$}}
\newcommand{\NTWO}{\mbox{$\widetilde{\chi}_2^0$}}
\newcommand{\NTHREE}{\mbox{$\widetilde{\chi}_3^0$}}
\newcommand{\NFOUR}{\mbox{$\widetilde{\chi}_4^0$}}
\newcommand{\CONE}{\mbox{$\widetilde{\chi}_1^{\pm}$}}
\newcommand{\CTWO}{\mbox{$\widetilde{\chi}_2^{\pm}$}}
\newcommand{\CONEP}{\mbox{$\widetilde{\chi}_1^{+}$}}
\newcommand{\CONEM}{\mbox{$\widetilde{\chi}_1^{-}$}}
\newcommand{\none}{\NONE}
\newcommand{\ntwo}{\NTWO}
\newcommand{\cone}{\CONE}
\newcommand{\coneb}{\CONEB}
\newcommand{\NTGNO}{\mbox{$\NTWO \rightarrow \gamma \NONE$}}
\newcommand{\NTGTGR}{\mbox{$\NONE \rightarrow \gamma \Gravitino$}}
\newcommand{\tanbeta}{\tan\beta}
\newcommand{\bfTeV}{\ensuremath{\bf{Te\kern -0.1em V}}\xspace}
\newcommand{\rrr}{\rightarrow}
\newcommand{\RRR}{\Rightarrow}
\newcommand{\Etgamma}{\ensuremath{\mathrm{E_T^{\gamma}}}}
\newcommand{\Etjet}{\ensuremath{\mathrm{E_T^{jet}}}}
\newcommand{\Ptjet}{\ensuremath{\mathrm{P_T^{jet}}}}
\newcommand{\Etlepton}{\ensuremath{\mathrm{E_T^{\ell}}}}
\newcommand{\Etelectron}{\ensuremath{\mathrm{E_T^{e}}}}
\newcommand{\Eelectron}{\ensuremath{\mathrm{E_T^{e}}}}
\newcommand{\Ptelectron}{\ensuremath{\mathrm{P_T^{e}}}}
\newcommand{\Ptlepton}{\ensuremath{\mathrm{P_T^{\ell}}}}
\newcommand{\Ptmuon}{\ensuremath{\mathrm{P_T^{\mu}}}}
\newcommand{\Etmuon}{\ensuremath{\mathrm{E_T^{\mu}}}}
\newcommand{\Emuon}{\ensuremath{\mathrm{E^{\mu}}}}
\newcommand{\Pttau}{\ensuremath{\mathrm{P_T^{\tau}}}}
\newcommand{\Ettau}{\ensuremath{\mathrm{E_T^{\tau}}}}
\newcommand{\LPX}{Lepton + Photon + X~}
\newcommand{\lpX}{lepton + photon + X~}
\def\stilde{\widetilde}

%% file: Loginov/loginov.tex
\subsection{Introduction}

The top quark was discovered almost twenty years ago. However,
couplings of the top quark to the neutral electroweak (EW) gauge
bosons ($\gamma$ and $Z$) have not yet been directly measured. Due to
the large mass, the top quark may play a special role in EW symmetry
breaking (EWSB), and therefore new physics connected with EWSB can
manifest itself in top precision observables. Possible signals for new
physics are deviations of the $\ttbar\gamma$, $\ttbar Z$ (and also
$tbW$) couplings from the values predicted by the Standard Model (SM).

While the associated $\ttbar W$ production cross section measurement
is an important test of the Standard Model predictions for this low
cross section process, it has little to do with the top quark (see
Fig.~\ref{f:ttv_diagrams}). The $tbW$ coupling has been constrained
via measurements of the single top quark production cross section (see
Section~\ref{s:singletop}) as well as via top quark width
measurements, and hence it is not covered in this Section.

The $\ttbar \gamma$ and $\ttbar Z$ couplings can not be constrained
via measurements of $\ttbar$ production at hadron colliders via
intermediate virtual $\gamma$ and $Z$ bosons, since the cross section
for $pp \rightarrow \ttbar$ is dominated by processes involving QCD
couplings. Unlike at a linear $e^+e^-$ collider, the LHC's capability
of associated $\ttbar\gamma$ and $\ttbar Z$ production has the
advantage that the $\ttbar\gamma$ and $\ttbar Z$ couplings are not
entangled. The $\ttbar \gamma$ and $\ttbar Z$ couplings may be
measured via analysis of direct production of $\ttbar$ pairs in
association with a $\gamma$ or $Z$ boson, respectively.

\subsection{$\ttbar \gamma$}

\begin{figure}[htbp]
  \begin{center}
    \mbox{
`\subfigure[Radiative top quark production]{\includegraphics[width=0.152\textwidth]{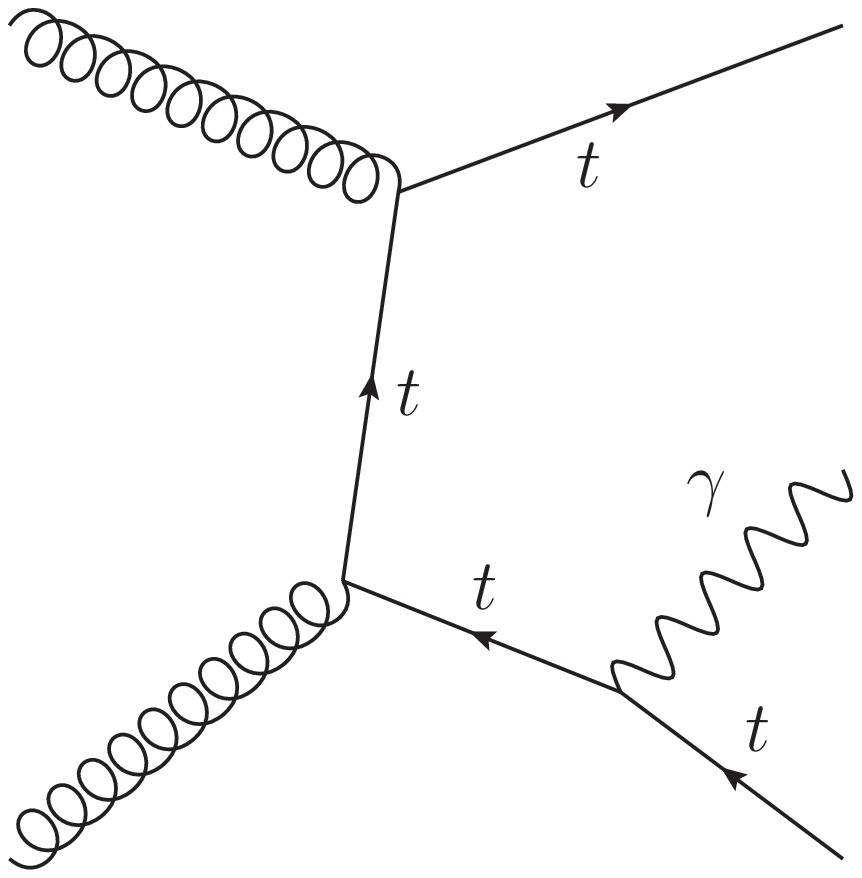}\hspace{0.10cm}\includegraphics[width=0.152\textwidth]{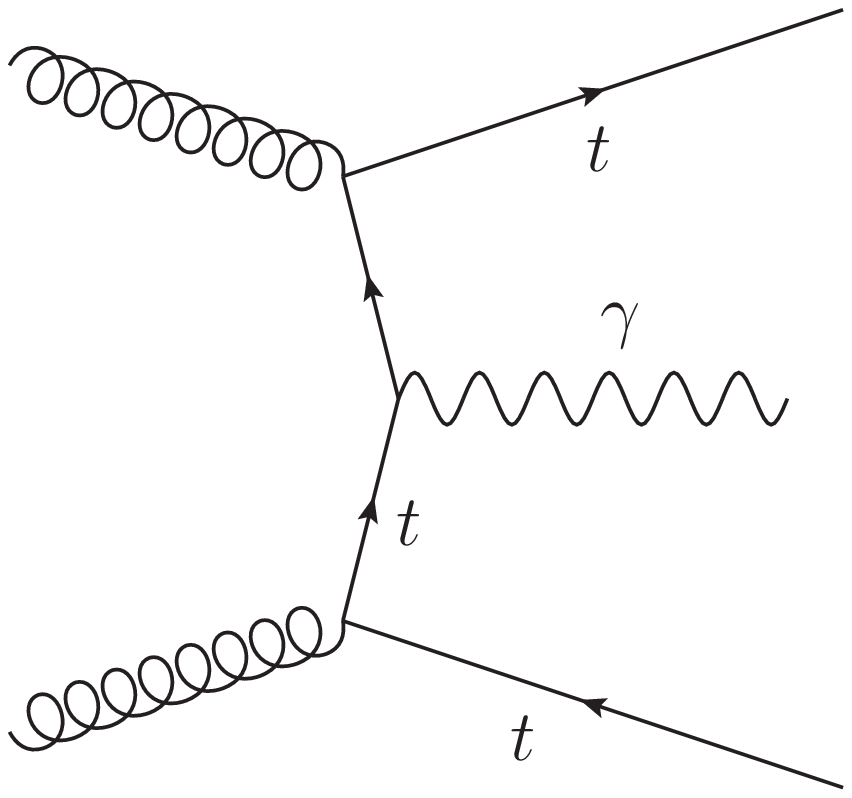}\hspace{0.10cm}\includegraphics[width=0.152\textwidth]{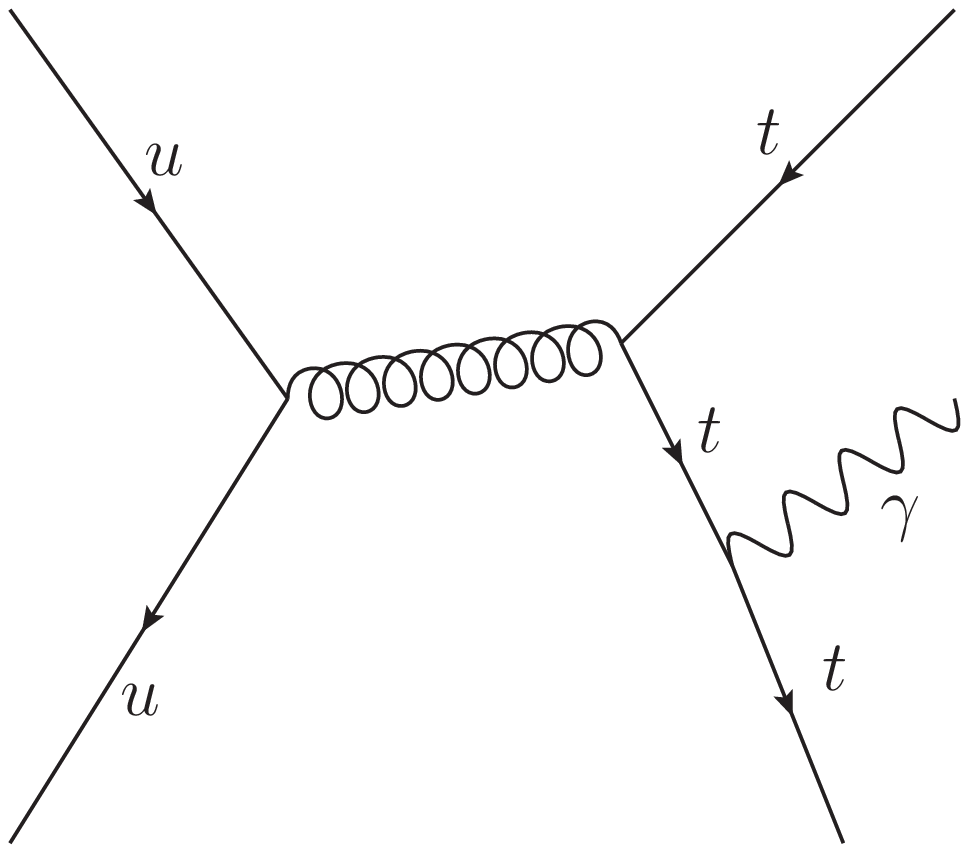}}
      }
    \mbox{
\subfigure[Radiative top quark decay]{\includegraphics[width=0.162\textwidth]{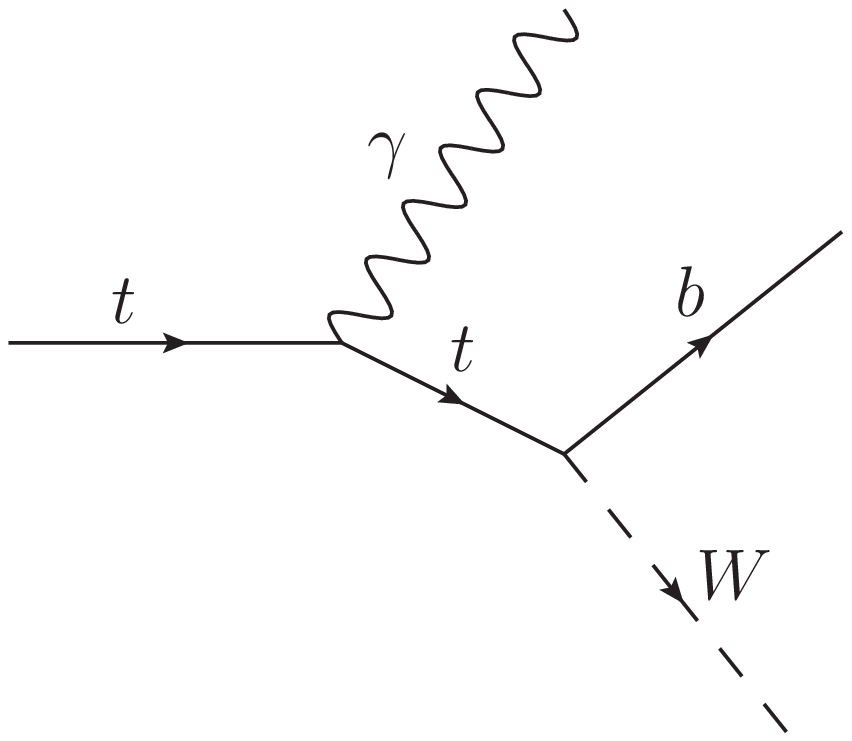}\includegraphics[width=0.162\textwidth]{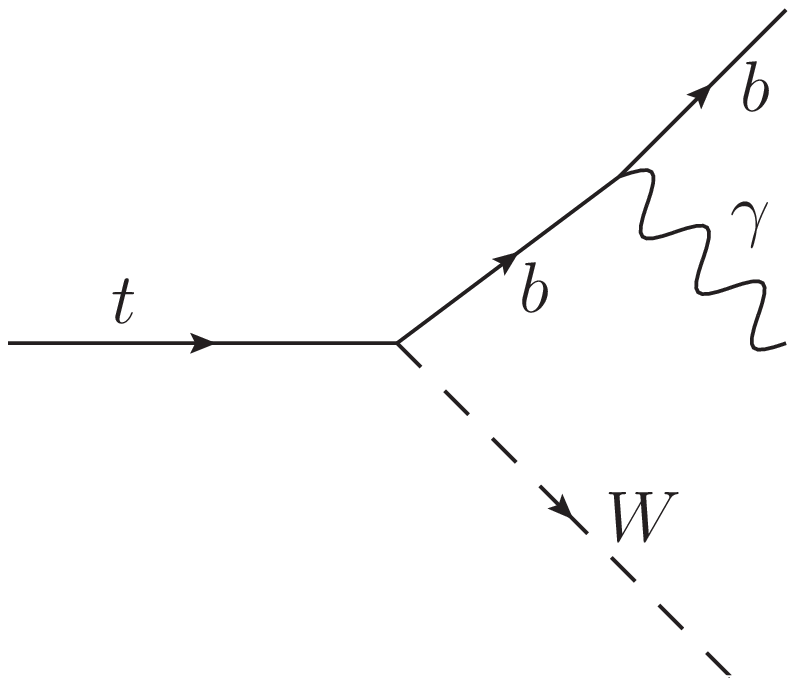}\includegraphics[width=0.162\textwidth]{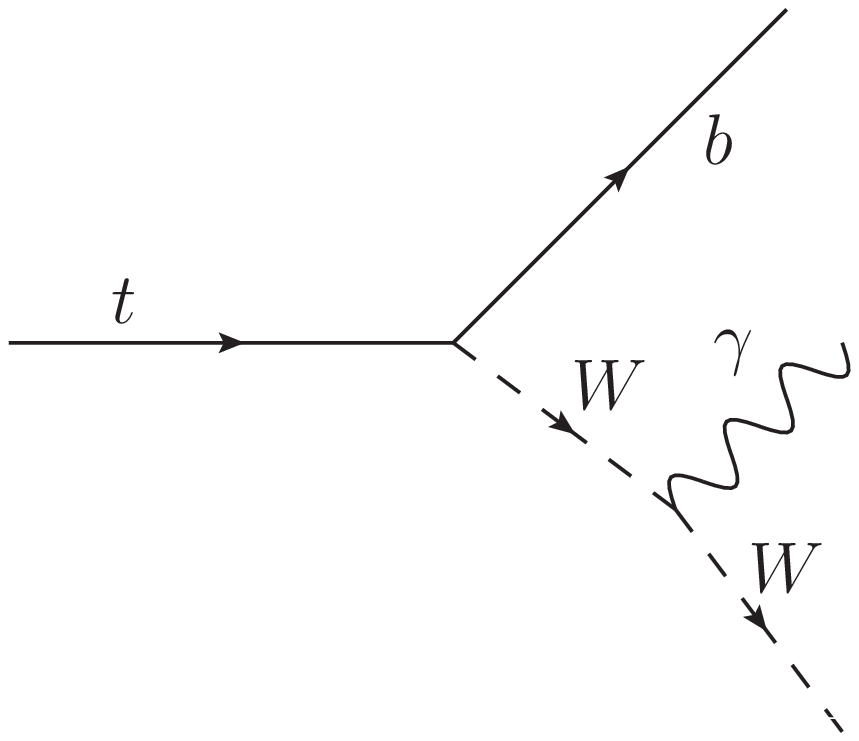}}
      }
    \caption {Representative Feynman diagrams for the $\ttg$
      process. Radiative top quark decay diagrams also include
      radiation from decay products of the $W$.}
    \label{f:ttgamma_diagrams}
  \end{center}
\end{figure}

Radiative $\ttbar$ production, $\ttbar\gamma$, can be classified into
the radiative top quark production and radiative top quark decay, as
shown in Fig.~\ref{f:ttgamma_diagrams}. The coupling of the top quark
to photons (and therefore the $\ttbar\gamma$ production cross section)
is sensitive to the electric charge of the top quark. The charge of
the top quark has been measured via its decay products using the track
charge method (a weighted sum of the charges of tracks associated with
the $b$-jet) and the soft lepton method (charge of the muon produced
in the semi-leptonic decay $b\rightarrow\mu\nu+X$ is defined by the
$b$-quark charge) at both the Tevatron and LHC. The $\ttbar \gamma$
can also yield constraints to excited top quarks ($t\rightarrow
t^*\gamma$) production. In addition, the $\ttbar\gamma$ signature is
an important control sample for $\ttbar Z$ and $\ttbar
H,~H\rightarrow\gamma\gamma$ analyses.

The distinction between the two classes of processes is indispensable
for determination of couplings of the top quark to the photon as
discussed below. However, the analyses performed so far by CDF and
ATLAS collaborations didn't measure the couplings, providing instead
measurements of the $\ttbar\gamma$ cross section, $\sigma_{\ttg}$. In
both the CDF and ATLAS $\ttbar\gamma$ analyses the strategy is to
perform data-driven background estimates for jets misidentified as
electrons, and for jets and electrons misidentified as photons. In
addition, the $\ttbar$ and $W\gamma$ production processes are used as
control samples for the $\ttbar$ part of the event and for the
photons, respectively. $Z\rightarrow e^+e^-$ with one of the electrons
misidentified as a photon is used to estimate the $e\rightarrow\gamma$
fake rate both in ATLAS and CDF. Both ATLAS and CDF use the
$Z\rightarrow e^+e^-$ process for real photons template, assuming
similarity between electrons and photons in the electromagnetic
calorimeter. CDF also uses $Z\gamma$ together with $W\gamma$ for
further photon identification optimization.

The CDF collaboration published the first experimental evidence for
$\ttg$ production by measuring the $\sigma_{\ttg}$ and the ratio
$R=\sigma_{\ttg}/\sigma_{\ttbar}$ with data corresponding to
6.0~fb$^{-1}$ of $\ppbar$ collisions at $\sqrt{s}=1.96$ TeV at the
Tevatron. The $\ttg$ cross section was measured to be $\sigma_{\ttg} =
0.18 \pm 0.08$~pb with a significance of
3.0~$\sigma$~\cite{cdf_ttgamma} for a photon $p_T > 6$ GeV (using a
photon $p_T > 10$ GeV in the reconstruction). The ratio was measured
to be $R = 0.024 \pm 0.009$ in a good agreement with the SM
expectation of $0.024 \pm 0.005$ for a charge 2/3 top quark. The
dominant uncertainty in these measurements arises from the finite
statistics of $\ttbar\gamma$ candidate events. However, the
$R=\sigma_{\ttg}/\sigma_{\ttbar}$ measurement is more precise than the
$\sigma_{\ttg}$ measurement due to the cancellation of the
$\ttbar$-related systematic uncertainties.

The ATLAS collaboration has performed the first $\ttbar\gamma$
production~\cite{atlas_ttgamma} cross section measurement at the LHC
in 1.04 fb$^{-1}$ of data at $\sqrt{s}=7$ TeV,
$\sigma_{\ttbar\gamma}=2.0\pm 0.5~(stat.)\pm 0.7~(syst.)\pm
0.08~(lumi.)$~pb for a photon $p_T$ threshold of 8 GeV (using a photon
$p_T > 15$ GeV in the reconstruction). The significance of the
measurement is 2.7~$\sigma$, and the measured cross section is in good
agreement with the SM expectation of $2.1\pm 0.4$~pb for a charge 2/3
top quark. The uncertainty of the measurement is dominated by the
photon identification efficiency, the initial and final state
radiation modeling and the jet energy scale systematics. The analysis
is being updated with the full 7 TeV dataset (systematic uncertainties
are also expected to get reduced).

The LO $\ttbar\gamma$ production cross section increases by a factor
of 5 from 7 TeV to 14 TeV center-of-mass energy for photons with
$p_T>20$~GeV~\cite{Stelzer:1994ta,Maltoni:2002qb}. With the much
larger statistics expected after the upcoming 2013-2014 shutdown,
analysis of the top quark--photon couplings is possible.

Furthermore, to isolate events with photon emission from top quarks,
the photon radiation from the $W$ and its decay products, as well as
from the $b$ quarks and from the initial-state quarks should be
suppressed, as detailed in Ref.~\cite{tt_couplings_baur}. To suppress
photon radiation from the $b$ quarks (leptons) a large $\Delta
R(\gamma,b)$ ($\Delta R(\gamma,\ell)$) is required. Photon emission
from $W$ decay products can be eliminated by requiring that the
invariant mass of the $jj\gamma$ system $m(jj\gamma) > 90~{\rm GeV}$
and the $\ell\gamma\met$ cluster transverse mass $m_T(\ell\gamma,\met
> 90~{\rm GeV}$. The radiation from the initial state quarks is hard
to suppress without simultaneously reducing the signal cross section.
However, as at the LHC $\ttbar$ production by gluon fusion dominates,
it is not an issue (unlike at the Tevatron). In addition, the event is
required to be consistent either with $\ttbar\gamma$ production, or
with the $\ttbar$ production with radiative top decay by performing
top quark and top quark + photon mass reconstruction as shown in
Ref.~\cite{tt_couplings_baur}.

Isolating events with photon emission from top quarks, as well as
performing a ratio $R=\sigma_{\ttg}/\sigma_{\ttbar}$ measurement to
reduce $\ttbar$-related systematic uncertainties is the strategy for
future $\ttbar\gamma$ analyses. At the LHC, with 300 $\invfb$ several
thousand signal events are expected, therefore precise determination
of the $\ttbar\gamma$ couplings is possible~\cite{tt_couplings_baur}
using the lepton plus jets channel. In addition, a precise $\ttbar\gamma$
couplings measurement can be performed using the dilepton channel
which should provide a smaller systematic uncertainty (due to fewer
jets in the event) but a slightly larger statistical uncertainty
compared to the lepton plus jets channel. With 3000 $\invfb$,
differential measurements of $\ttbar\gamma$ couplings (for instance, as a
function of photon $p_T$) as well as differential $\ttbar\gamma$ cross
section measurements, can be performed.

\subsection{$\ttbar Z$}

\begin{figure}[!htbp]
  \begin{center}
    \mbox{
\subfigure[$\ttbar Z$ process]{\includegraphics[bb = 103 110 545 275,clip=,width=0.45\textwidth]{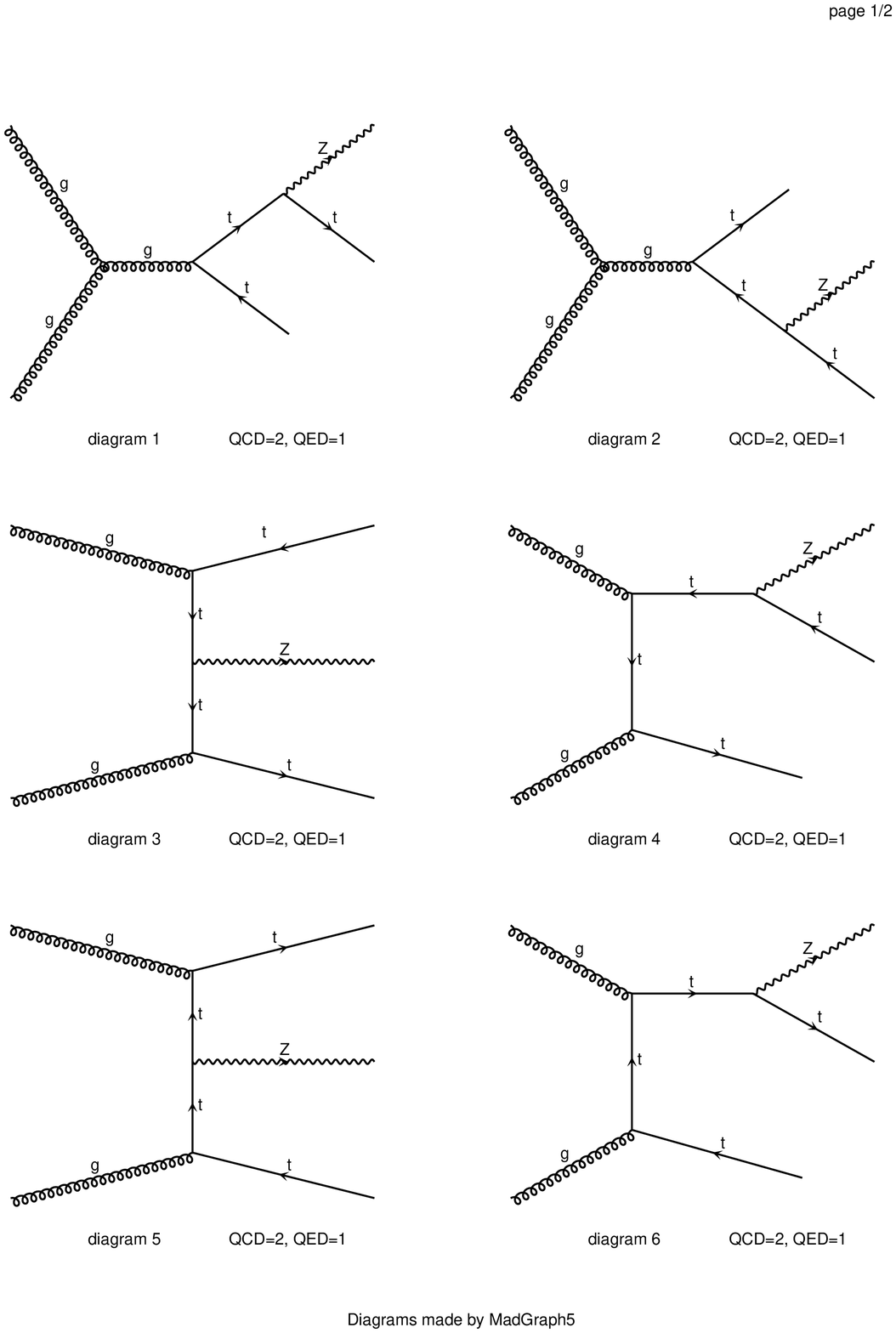}}\hspace{0.5cm}
      }
    \mbox{
\subfigure[$\ttbar W$ process]{\includegraphics[bb = 83 555 540 715,clip=,width=0.45\textwidth]{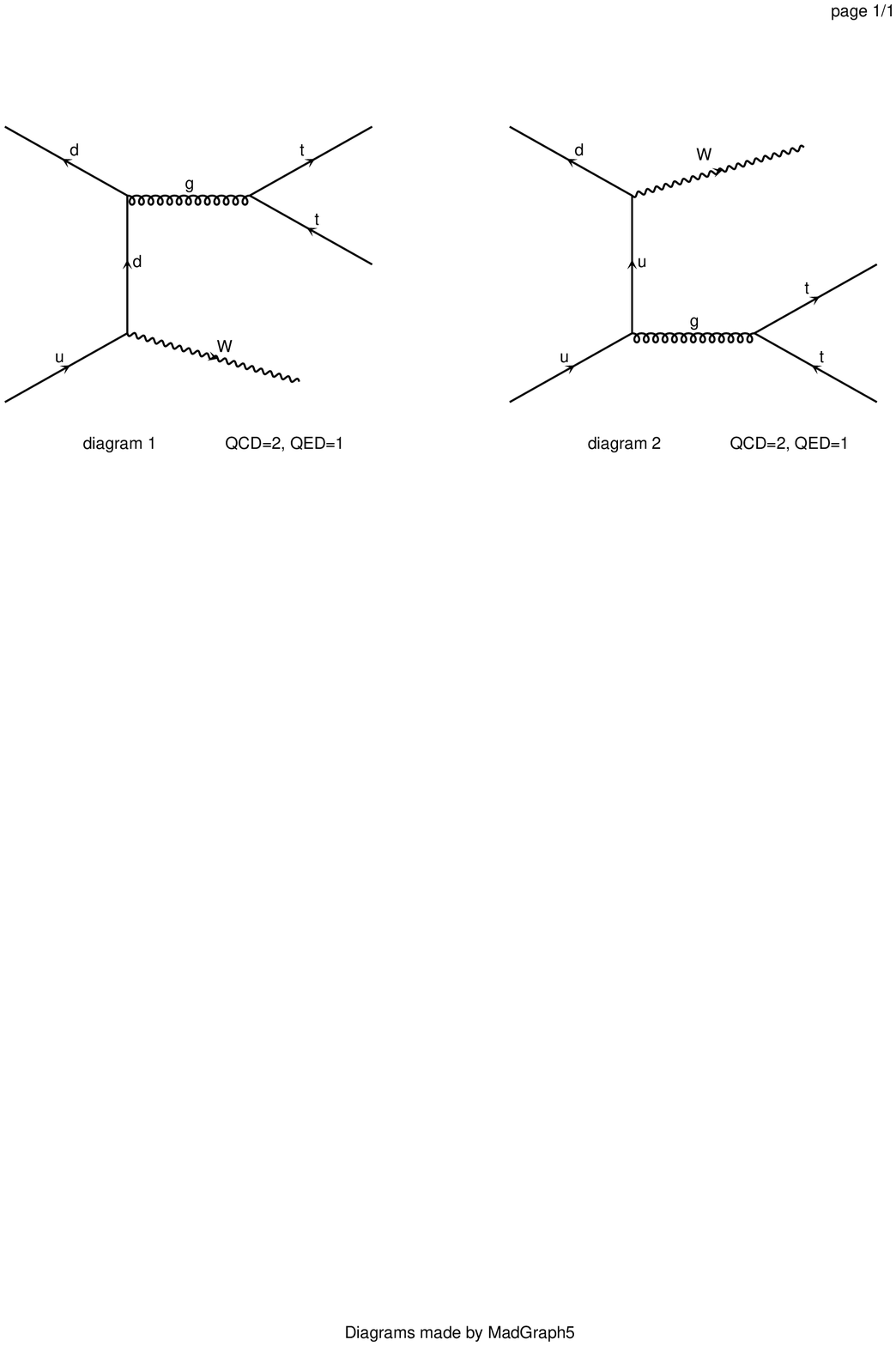}}
      }
    \caption {
Representative Feynman diagrams for the $\ttbar V$ process. }
    \label{f:ttv_diagrams}
  \end{center}
\end{figure}

The associated $\ttbar Z$ production is directly sensitive to $\ttbar
Z$ couplings. Representative Feynman diagrams for $\ttbar V$ (in this
Section $V$ = $Z$, $W$) production are shown in
Fig.~\ref{f:ttv_diagrams}.

The associated production of $\ttbar$ and $Z$ or $W$ bosons has been
measured by CMS in 5 fb$^{-1}$ of pp collisions at
$\sqrt{s}=7~$TeV~\cite{Chatrchyan:2013qca}. The measurements exploit
the fact that SM events with two prompt same-sign isolated leptons in
the final state, as well as trilepton events, are very rare. 
%, as shown in Fig.~\ref{fig:yields}. 
A data-driven estimation procedure is employed to estimate the
background contribution of jets misidentified as leptons.

%\begin{figure}[t]
%\centering
%\includegraphics[width=0.38\textwidth]{\home/Loginov/channelYields}
%\includegraphics[width=0.38\textwidth]{\home/Loginov/ObsPred_TTbarWSel}
%\caption{Event yields in $\ttbar V$ analysis~\cite{Chatrchyan:2013qca}
% for the trilepton (left) and same-sign dilepton (right) selections.}
%\label{fig:yields}
%\end{figure}

\begin{figure}[t]
\centering
\includegraphics[width=0.5\textwidth]{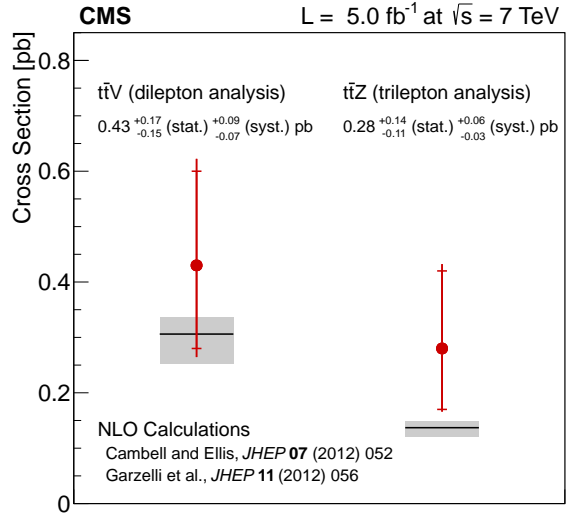}
\caption{Measurements of the $\ttbar Z$ and $\ttbar V$ production
  cross sections~\cite{Chatrchyan:2013qca}, in the same-sign
  dilepton~(left) and trilepton channel~(right), respectively. The
  measurements are compared to the next-to-leading order Standard
  Model calculation (horizontal black lines) and their uncertainty
  (grey bands). Internal error bars for the measurements represent the
  statistical component of the uncertainty.  }
\label{fig:measurement_vs_NLO}
\end{figure}

A direct measurement of the $\ttbar Z$ cross section $\sigma_{\ttbar
  Z} = 0.28^{+0.14}_{-0.11} (stat.)^{+0.06}_{-0.03} (syst.)$~pb is
obtained in the {\em trilepton channel}, observing 9 events with the
expected background of 3.2 $\pm$ 0.8 events. The signal significance
is 3.3 standard deviations from the background hypothesis. In the {\em
  dilepton channel} a total of 16 events is selected in the data,
compared to an expected background contribution of $9.2\pm 2.6$
events. The presence of a $\ttbar V$ ($V=W, Z$) signal is established
with a significance of 3.0 standard deviations. The $\ttbar V$ process
cross section is measured to be $\sigma_{\ttbar V} =
0.43^{+0.17}_{-0.15} (stat.) ^{+0.09}_{-0.07} (syst.)$~pb. Both
$\sigma_{\ttbar V}$ and $\sigma_{\ttbar Z}$ cross section measurements
are compatible with the NLO predictions. These results are shown in
Fig.~\ref{fig:measurement_vs_NLO} together with the next-to-leading
order SM predictions.

%\begin{figure}[!htbp]
%\centerline{\epsfxsize=3.8in\epsfbox{\home/Loginov/ttV_cms.eps}}
%\caption[*]{ Measurement of the $\ttbar V$ production cross
%section~\cite{Chatrchyan:2013qca}: the cross section from the trilepton channel (blue
%triangle), from the same sign dilepton channel (grey square) and the
%combination of the two measurements (black circle) are compared to
%the NLO calculation (red line). Internal error bars for the
%measurements represent the statistical component of the uncertainty.}
%\label{fig:ttv}
%\end{figure}

The ATLAS collaboration also performed $\ttbar Z$
analysis~\cite{ttz_atlas}, using a much tighter selection to suppress
backgrounds, that resulted in observing one event in data and setting
a limit $\sigma_{\ttbar Z} < 0.71$~pb at 95\% CL consistent with the
CMS measurement and with theoretical predictions~\cite{ttz_baur}.

The uncertainties of the $\ttbar Z$ measurements up to date are
dominated by statistics. The LO $\ttbar Z$ production cross section
increases by a factor of $\sim$ 1.4 from 7 TeV to 8 TeV center-of-mass
energy~\cite{Stelzer:1994ta,Maltoni:2002qb}, so the expected decrease
in the statistical uncertainty with the dataset collected in 2012 is a
factor of 2.5. Therefore, statistical uncertainties are expected to
dominate the $\sigma_{\ttbar V}$ and $\sigma_{\ttbar Z}$ measurements
performed with 2012 data.

However, the LO $\ttbar Z$ production cross section increases by
roughly an order of magnitude from 7 TeV to 14 TeV center-of-mass
energy~\cite{Stelzer:1994ta,Maltoni:2002qb}, therefore precise
measurements of the $\ttbar Z$ production cross section can be
performed after the 2013-2014 shutdown.  According to
Ref.~\cite{tt_couplings_baur}, with 300 $\invfb$ of 14 TeV collisions
data, the $\ttbar Z$ vector (axial vector) coupling can be determined
with an uncertainty of 45 -- 85\% (15 -- 20\%), whereas the
dimension-five dipole form factors can be measured with a precision of
50 -- 55\%.  For 3000 $\invfb$ of data expected at the High Luminosity
(HL) LHC, the limits are expected to improve by a factor of 3.

%% file: Vos/vos.tex
\subsection{Linear Collider prospects\protect\footnote{Author: Marcel Vos, IFIC (UVEG/CSIC)}}

A future $e^+e^-$ collider~\cite{ilc,clic} 
operated at center-of-mass energy greater than 
the top quark pair production threshold ($\sqrt{s} > 2 m_t$) offers 
excellent prospects for a precision top physics programme. 

One of the key assets of lepton collider is calculability; many observable 
quantities can be predicted at the per mil level. 
The $t\bar{t}$ production rate can be calculated to excellent precision. 
For $\sqrt{s} =$ 500~\gev{} 
QCD corrections have been determined to order $\alpha^3_s$~\cite{Kiyo:2009gb}. 
The $N^4LO$ contribution to the rate is estimated from scale variations to 
be approximately 3 per mil. 

Top quark production at an $e^+e^-$ collider is less copious than at the LHC. 
At the LC $e^+e^- \rightarrow Z/\gamma^* \rightarrow t \bar{t}$ 
is the dominant top quark production process. All $t\bar{t}$ final states 
are readily isolated, with background levels due to other six-fermion 
processes below 5\%.
At $\sqrt{s}=$ 500~\gev{} the (unpolarized) cross-section is 
approximately 0.6 pb. At that energy an integrated 
luminosity of 125\mbox{pb$^{-1}$} is envisaged to be accumulated 
each year. 
A four year period thus yields a sample of several 100.000 top quark pairs,
shared equally between two beam polarization configurations 
($\cal{P}_e^{-},\cal{P}_e^{+} = \pm$ 80\%,$\mp$ 30\%).

%The forward-backward asymmetry in $t\bar{t}$ production has 
%been calculated to $NNLO$~\cite{Bernreuther:2006vp} and the $N^3LO$ 
%contribution is less than 1\%.
%One-loop electroweak corrections have been found to be 
%sizeable~\cite{Fleischer:2003kk, Khiem:2012bp (several
%\% on the total rate, order 10\% on $A_{FB}$). A 
%two-loop result is required to perform the measurements to the 
%required precision.

The LC potential for the characterization of the $t \bar{t} Z$ and 
$t \bar{t} \gamma$ vertices was established using parton-level
studies a long time ago~\cite{AguilarSaavedra:2001rg}. For
compatibility with these studies and the LHC prospects of
Reference~\cite{tt_couplings_baur} we express the sensitivity in
terms of the form factors that characterize the current at the
$t \bar{t} Z$ and $t \bar{t} \gamma$ vertices:
\begin{equation}
\label{eq:snow}
\Gamma_\mu^{ttX}(k^2,\,q,\,\bar{q}) = ie \left\{
  \gamma_\mu \, \left(  \widetilde{F}_{1V}^X(k^2)
                      + \gamma_5\widetilde{F}_{1A}^X(k^2) \right)
+ \frac{(q-\bar{q})_\mu}{2m_t}
    \left(  \widetilde{F}_{2V}^X(k^2)
          + \gamma_5\widetilde F_{2A}^X(k^2) \right)
\right\} 
\end{equation}

Recently, the LC prospect were
revisited~\cite{Asner:2013hla,Amjad:2013tlv} with a detailed
simulation of the detector response, including the impact of $\gamma
\gamma \rightarrow$ {\em hadrons} background. The analysis aims for
simultaneous determination of the photon and $Z$ boson form factors.
Three observables - the cross section, the forward-backward asymmetry
and the top quark polarization - are evaluated for two polarization
states of the electron and positron beams. These six independent
measurements are used to extract five CP conserving form factors:
$\widetilde F^{\gamma,Z}_{1V}$ and $\widetilde F^{\gamma,Z}_{2V}$ and
$\widetilde F^Z_{1A}$ ($F^\gamma_{1A}$ is taken to be 0 to preserve
gauge invariance).

After accounting for the selection efficiency the statistics of the sample
is sufficient to reduce the statistical uncertainty of these measurements 
to the per mil level. Systematic errors due to 
the reconstruction of the complex $t \bar{t}$ events are accounted for. 
The impact of a number of several other potential sources - such as the
expected error on the beam energy, the luminosity and the beam polarization
- is evaluated and found to be sub-dominant.

\begin{figure}[htbp]
  \begin{center}
 \includegraphics[width=0.4\textwidth]{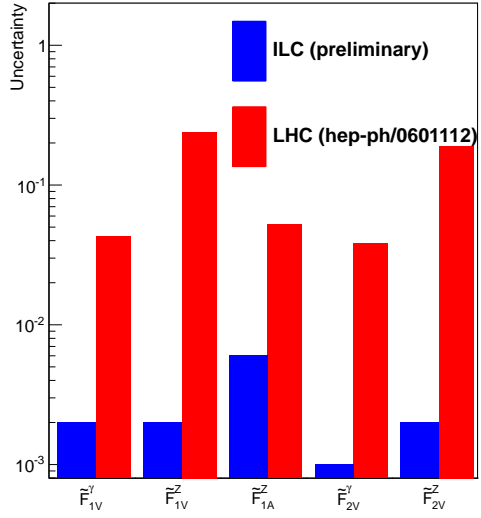}
    \caption {Sensitivity at $68.3\%$ CL for $CP$ conserving form factors $\widetilde F^X_{1V,A}$ and $\widetilde F^X_{2V}$ defined in Eq.~\ref{eq:snow} at the LHC and at linear $\epem$ colliders.}
    \label{f:ilc_couplings}
  \end{center}
\end{figure}

The results are compared to the LHC prospects from 
Reference~\cite{Juste:2006sv} 
(assuming an integrated luminosity of 300~\mbox{fb$^{-1}$} 
at 14~\tev{}) in Figure~\ref{f:ilc_couplings}
and Table~\ref{tab:ilc_couplings}.
The estimated sensitivity of the LC is an order of magnitude greater for 
all couplings than that expected at the LHC, and exceeds it by two orders of
magnitude for the vectorial couplings of the $Z$ boson.

\begin{table}[t]
\begin{center}
\begin{footnotesize}
%\begin{tabular}{ccc|ccc}
\begin{tabular}{|ccccc|}
\hline 
 Coupling & SM value & LHC~\protect\cite{Juste:2006sv}  & $e^+e^-$~\protect\cite{AguilarSaavedra:2001rg}& 
           $e^+e^-$~\protect\cite{Asner:2013hla,Amjad:2013tlv}\\
& & $\mathcal{L}=300~\invfb$ &  $\mathcal{L}=300~\invfb$  & $\mathcal{L}=500~\invfb$\\
& &                                                &     ${\cal P},{\cal P}'=-0.8,0$                                        & ${\cal P},{\cal P}' =\pm0.8,\mp0.3$\\
\hline
$\Delta\widetilde F^\gamma_{1V}$ & 
0.66 &
$\begin{matrix} +0.043 \\[-4pt] -0.041\end{matrix}$ & 
$\begin{matrix} - \\[-4pt] - \end{matrix}$&   
%$\mathcal{L}=200~\invfb$& 
%$\Delta\widetilde F^Z_{1V}$ &
%$\begin{matrix} +0.24 \\[-4pt]  -0.62\end{matrix}$ & 
$\begin{matrix} +0.002 \\[-4pt] -0.002 \end{matrix}$
\\
%$\Delta\widetilde F^\gamma_{1A}$ & 
%$\begin{matrix} +0.051 \\[-4pt] -0.048\end{matrix}$ & 
%$\begin{matrix} +0.011 \\[-4pt] -0.011\end{matrix}$ , 
%    $\mathcal{L}=100~\invfb$ & 
%$\Delta\widetilde F^Z_{1A}$ &
%$\begin{matrix} +0.052 \\[-4pt]  -0.060\end{matrix}$ & 
%$\begin{matrix} +0.009 \\[-4pt] -0.009\end{matrix}$   \\
%$\Delta\widetilde F^\gamma_{2V}$ & $\begin{matrix} +0.038 \\[-4pt]
%-0.035\end{matrix}$ & $\begin{matrix} +0.038 \\[-4pt]
%-0.038\end{matrix}$ , 200~fb$^{-1}$  & $\Delta\widetilde F^Z_{2V}$ &
%$\begin{matrix} +0.27 \\[-4pt] 
%-0.19\end{matrix}$ & $\begin{matrix} +0.009 \\[-4pt]
%-0.009\end{matrix}$ , 200~fb$^{-1}$ 
%\\
%$\Delta\widetilde F^\gamma_{2A}$ & $\begin{matrix} +0.16 \\[-4pt]
%-0.17\end{matrix}$ & $\begin{matrix} +0.014 \\[-4pt]
%-0.014\end{matrix}$ , 100~fb$^{-1}$   & $\Delta\widetilde F^Z_{2A}$ &
%$\begin{matrix} +0.28 \\[-4pt] 
%-0.27\end{matrix}$ & $\begin{matrix} +0.052 \\[-4pt]
%-0.052\end{matrix}$ , 100~fb$^{-1}$  
%\\     
$\Delta\widetilde F^Z_{1V}$ &
0.23 &
$\begin{matrix} +0.240 \\[-4pt]  -0.620\end{matrix}$ & 
$\begin{matrix} +0.004 \\[-4pt] -0.004\end{matrix}$ &
%$\Delta\widetilde F^\gamma_{1V}$ & 
%$\begin{matrix} +0.043 \\[-4pt] -0.041\end{matrix}$ & 
$\begin{matrix} +0.002 \\[-4pt] -0.002\end{matrix}$  
\\
$\Delta\widetilde F^Z_{1A}$ &
-0.59 &
$\begin{matrix} +0.052 \\[-4pt]  -0.060\end{matrix}$ & 
$\begin{matrix} +0.009 \\[-4pt] -0.013\end{matrix}$ &
%$\Delta\widetilde F^\gamma_{1A}$ & 
%$\begin{matrix} +0.051 \\[-4pt] -0.048\end{matrix}$ & 
$\begin{matrix} +0.006 \\[-4pt] -0.006\end{matrix}$ 
\\
$\Delta\widetilde F^\gamma_{2V}$ & 
0.015 &
$\begin{matrix} +0.038 \\[-4pt] -0.035\end{matrix}$ & 
$\begin{matrix} +0.004 \\[-4pt] -0.004\end{matrix}$ &   
%$\Delta\widetilde F^Z_{2V}$ &
%$\begin{matrix} +0.27 \\[-4pt]  -0.19\end{matrix}$ & 
$\begin{matrix} +0.001 \\[-4pt] -0.001\end{matrix}$
\\
$\Delta\widetilde F^Z_{2V}$ & 
0.018 &
$\begin{matrix} +0.270 \\[-4pt] -0.190\end{matrix}$ & 
$\begin{matrix} +0.004 \\[-4pt] -0.004\end{matrix}$ & 
%$\Delta\widetilde F^Z_{2A}$ &
 $\begin{matrix} +0.002 \\[-4pt] -0.002\end{matrix}$  
\\
%$\Delta\widetilde F^\gamma_{2V}$ & 
%$\begin{matrix} +0.038 \\[-4pt] -0.035\end{matrix}$ & 
%$\begin{matrix} +0.038 \\[-4pt] -0.038\end{matrix}$ , 200~fb$^{-1}$  & 
%$\Delta\widetilde F^Z_{2V}$ &
%$\begin{matrix} +0.27 \\[-4pt]  -0.19\end{matrix}$ & 
%$\begin{matrix} +0.009 \\[-4pt] -0.009\end{matrix}$ , $\mathcal{L}=200~\invfb$  &
%n.a.
%\\
%$\Delta\widetilde F^\gamma_{2A}$ & 
%$\begin{matrix} +0.16 \\[-4pt] -0.17\end{matrix}$ & 
%$\begin{matrix} +0.014 \\[-4pt] -0.014\end{matrix}$ , 100~fb$^{-1}$   & 
%$\Delta\widetilde F^Z_{2A}$ &
%$\begin{matrix} +0.28 \\[-4pt]  -0.27\end{matrix}$ & 
%$\begin{matrix} +0.052 \\[-4pt] -0.052\end{matrix}$ ,  $\mathcal{L}=100~\invfb$  &
%n.a.
%\\
\hline
$\Delta {\mbox Re}\, \widetilde F_{2A}^{\gamma}$& - & $\begin{matrix} +0.17 \\[-4pt] -0.17\end{matrix}$ &$\begin{matrix} +0.007 \\[-4pt] -0.007\end{matrix}$ &  \\
$\Delta {\mbox Re}\, \widetilde F_{2A}^{Z}$& - & $\begin{matrix} +0.35 \\[-4pt] -0.35\end{matrix}$ &$\begin{matrix} +0.008 \\[-4pt] -0.008\end{matrix}$ &  \\
$\Delta {\mbox Im}\, \widetilde F_{2A}^{\gamma}$& - & $\begin{matrix} +0.17 \\[-4pt] -0.17\end{matrix}$  &$\begin{matrix} +0.008 \\[-4pt] -0.008\end{matrix}$ &  \\
$\Delta {\mbox Im}\, \widetilde F_{2A}^{Z}$& - & $\begin{matrix} +0.035 \\[-4pt] -0.035\end{matrix}$ & $\begin{matrix} +0.015 \\[-4pt] -0.015\end{matrix}$ &  \\
\hline
\end{tabular}
\end{footnotesize}
\caption{\sl Sensitivity at $68.3\%$ CL for $CP$ conserving form factors 
$\widetilde F^X_{1V,A}$ and $\widetilde F^X_{2V}$ and the top quark magnetic 
and electric dipole form factors $\widetilde F^V_{2A}$ of Eq.~\ref{eq:snow}.
LHC prospects are compared to two linear $\epem$ collider studies. 
The assumptions for the integrated luminosity and, for $\epem$ colliders, 
the beam polarisation, are indicated in the table header. 
}
\label{tab:ilc_couplings}
\end{center}
\end{table}

The control over the beam polarization proves
sufficient to disentangle photon and Z boson form factors. A simultaneous
extraction of $\widetilde F_1$ and $\widetilde F_2$, independent of 
any assumption on the other form factor, turns out to be more cumbersome. 
Therefore, the results in Figure~\ref{f:ilc_couplings} correspond to a 
simultaneous determination of the four $\widetilde F_1$ form factors, while 
the two $\widetilde F_2$ form factors are kept at their Standard Model 
values, and vice versa. 
The potential to disentangle $\widetilde F_1$ and $\widetilde F_2$ improves 
at larger center-of-mass energy. The study of 
Reference~\cite{AguilarSaavedra:2012vh} 
shows that the equivalent effective operators can be constrained 
simultaneously if measurements 
at $\sqrt{s} =$ 500~\gev{} are combined with measurements at 1~\tev{}.

The potential of the LC to uncover Beyond the Standard Model contributions
to the top quark electric and magnetic dipole moment has not yet been 
evaluated in full simulation. The estimates at parton-level from 
Reference~\cite{AguilarSaavedra:2001rg} are compared to the LHC prospects
in Table~\ref{tab:ilc_couplings}.

Summarizing: an $e^+e^-$ collider with $\sqrt{s} > 2 m_t$ offers a powerful
precision top physics programme. The sensitivity for anomalous 
electroweak couplings of the top quark is boosted by an order of magnitude
with respect to the expected LHC potential.

%% file: Adelman/adelman.tex
\providecommand{\met}   {\ensuremath{E_{\mathrm{T}}^{\mathrm{miss}}}}
\def\ttbar{\ensuremath{t\bar{t}}}
\def\bbbar{\ensuremath{b\bar{b}}}
\def\ccbar{\ensuremath{c\bar{c}}}
\def\mbb{\ensuremath{m_{\bbbar}}}
\def\ttH{\ensuremath{\ttbar H}}
\def\WH{\ensuremath{WH}}
\def\ZH{\ensuremath{ZH}}
\def\pt{\ensuremath{p_{\mathrm{T}}}}
\def\mt{\ensuremath{m_{\mathrm{T}}}}
With the decades-long search for the Higgs boson complete and a
resonance consistent with Standard Model (SM) Higgs boson production
observed~\cite{atlashiggs, cmshiggs, tevhiggs}, the era of precision
Higgs boson studies now begins.  One particularly important property
of the Higgs boson is its coupling to top quarks. By far the most
massive quark, as well as the most massive fundamental particle
observed to-date, the top quark has Yukawa couplings close to unity,
making (first observation and then) measurements of~\ttH~particularly
important.  $H\rightarrow \gamma\gamma$ measurements from both ATLAS
and CMS~\cite{atlasdiphoton,cmsdiphoton}, which show larger signal
strength than expected in the SM in a channel with non-zero
contributions from loops involving top quarks, make a measurement of
the top-Higgs coupling even more important.

Measurements of the~\ttH~final state are not trivial, as
the~\ttbar~system itself is already quite complicated. Due to small
cross sections for~\ttH~production when compared with non-associated Higgs
production (or even when compared with~\ZH~and~\WH~production), analyses
so far have only searched for~\ttH~production with subsequent
$H\rightarrow \bbbar$ decay, which gives the largest branching
fraction. The final state is thus $\ttbar + \bbbar$~with a resonance
in the~$\bbbar$~mass. The main irreducible background is
then~$\ttbar$~production in association with extra jets, typically of
heavy flavor.

\subsection{Searches for $\ttbar+H$}
There are four direct searches in the literature
for~\ttH~production. CMS sets the most stringent limits, with an
observed (expected) upper limit at 95\% confidence level (C.L.) of
4.6x (3.8x) the SM expectation~\cite{cmstth} using the full 2011 LHC
$\sqrt{s} = 7$~TeV data set (5 fb$^{-1}$). The analysis uses both the
lepton+jets and dilepton topologies of $\ttbar$ decay. ATLAS also uses
the full 2011 data set (4.7 fb$^{-1}$) and sets observed (expected)
upper limits of 13.1x (10.5x) the SM expectation~\cite{atlastth} at
95 \% C.L., using only the single lepton decay channel. CDF has two
analyses: the full Run II Tevatron data set (9.45 fb$^{-1}$) is used
in the single lepton topology, with corresponding observed (expected)
upper limits of 20.5x (12.6x) the SM expectation~\cite{cdftth} at
95 \% C.L. A second CDF analysis~\cite{cdftth2} examines 0-lepton
events to captures both all-hadronic $\ttbar$ decays and single lepton
decays where the lepton was not reconstructed. The analysis is unique,
but suffers from extremely large backgrounds (from QCD), and in 5.7
fb$^{-1}$ sets observed (expected) upper limits of 36.2x (26.2x) the
SM prediction at a Higgs boson mass of 125~GeV.

The CMS analysis begins with the separation of events into single
lepton and dilepton categories. Tight leptons are required to be above
30~GeV. The definition of Loose leptons has relaxed identification, isolation and
$|\eta|$ requirements, and also lowers the $\pt$ thresholds to 10
(15-20) GeV for muons (electrons). Single lepton events are required
to contain tight leptons, whereas dilepton events can contain one
loose and one tight lepton. The 3 leading jets are required to be
above 40~GeV, and any additional jets are required to have $\pt
>$~30~GeV. Events are further separated based on the number of jets
and tags: 4 jets (3 or $\ge$ 4 tags), 5 jets (3 or $\ge$ 4 tags) and 6
or more jets (2, 3 or $\ge$ 4 tags) in the single lepton category; and
2 jets (2 tags) and $\ge$ 3 jets ($\ge$ 3 tags) in the dilepton
category. Artificial neural networks (ANN) are trained separately in
each category to separate signal and background. Inputs to the ANNs
include shape information (sphericity and aplanarity), $b$-tag weights
(to separate out $\ttbar$+light flavor from $\ttbar$+production of
extra associated heavy flavor), several Fox-Wolfram moments, event
mass, jet $\pt$, $\Delta R$ between tagged jets, $\Delta R$ between
leptons and jets, and the number of jets. Major systematic
uncertainties considered include jet energy scale (JES), scale
uncertainties for $\ttbar$ production (evaluated by changing
factorization and renormalization scales up and down by factors of two
in Madgraph, leading to uncertainties of 0-20\%) and $b$-tag scale
factor (SF) uncertainties. When scales are changed, the variations are
treated as uncorrelated between \ttbar+light flavor
jets, \ttbar\bbbar~and~\ttbar\ccbar, and give uncertainties that vary
as a function of the number of jets. For the lepton+jets channel, the
$b$-tag SF uncertainty is the dominant systematic effect. For the
dilepton channel, the factorization scale is the dominant systematic
uncertainty.

The ATLAS analysis examines only single lepton events. Electrons
(muons) are required to be above 25 (20) GeV. Jets are required to
have \pt~above 25 GeV. In the electron channel, $\met > 30$ GeV and
the transverse mass (\mt) between the lepton and the \met~is also
required to be above 30 GeV. In the muon channel, $\met > 20$ GeV and
$\met + \mt > 60$ GeV. Events are categorized based on the number of
jets and number of $b$-tags. The five control regions are: 4 jets (0,
1 or $\ge$ 2 tags), 5 jets (2 tags) and 6 or more jets (2 tags). The four 
signal categories are: 5 jets (3 or $\ge$ 4 tags) and $\ge$ 6 jets (3
or $\ge$ 4 tags). The final observable in the $\ge$6-jet signal
regions are the invariant mass of the pair of jets (\mbb) not selected
to come from $\ttbar$ decay. A kinematic fit for the $\ttbar$
hypothesis is used to assign observed jets to partons. Transfer
functions are used to describe detector resolution for jets matched to
partons at the truth level.  The scalar sum of the jet \pt~is used as
an observable in the 5-jet signal region and in the control regions
both to gain statistical power and also to help constrain nuisance
parameters {\it in-situ} in the fit to data. Major systematic
uncertainties considered include jet energy scale (JES), $b$-tag SF
uncertainties, and scale uncertainties for $\ttbar$ production. Scale
uncertainties have two factorization scale uncertainties (evaluated by
varying the default factorization scale $Q^2
= \Sigma_{\mbox{partons}}(m^2+p^2)$ up and down by factors of two and
taking the larger difference from nominal, as well as by using an
alternative default scale $Q^2 = x_1x_2s$, which dominates). The
renormalization scale uncertainty is evaluated by varying the scale up
and down by factors of factors of two. When varying scale
uncertainties, the fraction of \ttbar+jets events with heavy flavor is
found to vary by $\pm 50\%$, which is taken as an additional
uncertainty.

The single lepton CDF measurement requires a single electron or muon above 18 GeV and $\met > 10-25$ GeV, depending on the lepton flavor and $|\eta|$. Jets are required to have $\pt > 20$~GeV. Events are categorized based on the number of jets (4,5,6), with each jet category further split into 5 different $b$-tag subcategories using two different tagging algorithms. ANNs are trained in each subcategory to separate signal and background. The 18 input variables include \met, jet $\pt$ values, event mass, $\Delta R$ between tagged jets, lepton-jet masses, and dijet masses. The 4-jet 2-tag category is used is used to validate the ANN. Dominant systematic uncertainties are $b$-tag SFs, JES uncertainties, and background normalizations. A 10\% uncertainty is assumed on the $\ttbar$ + jets normalization, and a 100\% uncertainty is assumed on the \ttbar\bbbar~normalization.

As mentioned above, large uncertainties appear in these searches from normalization of $\ttbar$ produced in association with extra heavy flavor jets, the dominant background in the signal-enriched regions. CMS has made the first measurement of the ratio of production cross sections: $\sigma(\ttbar\bbbar)/\sigma(\ttbar jj)$ in a given visible phase space~\cite{cmsratio}. The result is larger than predictions from MADGRAPH and POWHEG, though with large uncertainties. Longer term and with significantly higher luminosities, the top-Higgs coupling will be better measured in other channels. Two promising examples are the diphoton and dimuon decay channels, where the presence of the Higgs boson produced in association with \ttbar~is seen via a resonance. ATLAS studied these channels~\cite{atlases} with hypothetical 300 and 3000 fb$^{-1}$ data sets containing large pile up at $\sqrt{s}=14$TeV, and claims the ability to measure $\Delta(\Gamma_t/\Gamma_g)$ to better than 55\% (25\%) with 300 (3000) fb$^{-1}$.

%% file: Khanov/khanov.tex
\renewcommand{\ttbar}{\mbox{$t\bar{t}$}}
\renewcommand{\bbbar}{\mbox{$b\bar{b}$}}
A measurement of \ttbar\ production with additional jets is an important test
of perturbative QCD at the LHC energy scale. The production of \ttbar+jets is
sensitive to initial and final state radiation as well as to other details of
the \ttbar\ production modelling. As such, it can serve as verification of
theoretical models and help to tune Monte Carlo generators.
Studies of \ttbar+jets may reduce systematic uncertainty due to
additional quark/gluon production, which is important for many
searches for new physics.

The first \ttbar+jets measurement was performed at the Tevatron by
the CDF collaboration~\cite{CDF-ttj}. At the LHC, both ATLAS and CMS
experiments studied the \ttbar+jets production in detail. The studies
performed at the LHC include measurement of \ttbar\ production with a veto
on additional central jet activity~\cite{ATLAS:2012al,CMS-PAS-TOP-12-023};
study of jet multiplicity in \ttbar\ events~\cite{ATLAS-CONF-2012-155,CMS-PAS-TOP-12-018,CMS-PAS-TOP-12-023};
the \ttbar+jets cross section measurement~\cite{ATLAS-CONF-2012-083};
and measurement of heavy flavor composition of \ttbar\ events~\cite{CMS-PAS-TOP-12-024}.

At hadron colliders, top quarks are predominantly produced in pairs.
In the Standard Model, a top quark decays almost 100\% of the time
to a $W$ boson and a $b$ quark. The $W$ boson can further decay leptonically
(to a lepton and a neutrino) or hadronically (giving rise to a pair of jets).
Consequently, there are three possible \ttbar\ final states:
both $W$'s decay leptonically (dilepton channel), one $W$
decays leptonically and the other one hadronically (lepton+jets channel), or
both $W$'s decay hadronically (all-hadronic channel). All analyses
described below are looking at top quark pairs produced in either
dilepton or lepton+jets channel.

\subsection{\ttbar\ with central jet veto}
ATLAS measured~\cite{ATLAS:2012al}
the fraction of \ttbar\ events produced without additional
jet(s) in a certain rapidity interval\footnote{
Rapidity is defined as $y=0.5\ln[(E+p_z)(E-p_z)]$
where $E$ is the energy and $p_z$ is the component of the momentum along
the beam direction.
}.
The measured quantity $f$ (gap fraction) is defined as $f(x)=n(x)/N$,
where $N$ is the total number of selected \ttbar\ events, and $n(x)$ is
the number of selected \ttbar\ events with additional jet veto.
The results were obtained for two veto definitions:
$x=Q_0$ (no additional jets with transverse momentum $p_T$ above
threshold $Q_0$ in a certain rapidity interval), and
$x=Q_{\rm sum}$ (the scalar transverse momentum sum of the additional jets
in the rapidity interval is less than $Q_{\rm sum}$).
By measuring the ratio of production rates, many systematic uncertainties cancel
out, which makes the result more sensitive to Monte Carlo modelling parameters.

The gap fraction
distributions as functions of $x$ (with a minimum $x$ value of 25~GeV)
were obtained for four jet rapidity intervals:
$|y|<0.8$, $0.8\le|y|<1.5$, $1.5\le|y|<2.1$, and for the full rapidity range, $|y|<2.1$.
The measurement was performed in the dilepton channel at a center-of-mass
energy of 7~TeV, using 2.05~fb$^{-1}$ of integrated luminosity.
The obtained gap fraction was corrected to the particle level using
a correction factor defined as the ratio of
the particle level gap fraction to the reconstructed gap fraction.

The results were compared to theoretical models implemented in
{\sc MC@NLO}~\cite{Frixione:2002ik,Frixione:2003ei} interfaced to {\sc Herwig}~\cite{Corcella:2000bw} and {\sc Jimmy}~\cite{Butterworth:1996zw},
{\sc Powheg}~\cite{Nason:2004rx,Frixione:2007vw} interfaced to either {\sc Pythia}~\cite{Sjostrand:2006za} or {\sc Herwig}/{\sc Jimmy},
{\sc Alpgen}~\cite{Mangano:2002ea} interfaced to {\sc Herwig}/{\sc Jimmy}, and
{\sc Sherpa}~\cite{Gleisberg:2008ta}.
It was found that while all models describe the data reasonably well within
the full $|y|<2.1$ veto interval, they tend to predict too much
jet activity in the most forward $1.5\le|y|<2.1$ region. In addition,
{\sc MC@NLO} underestimates the data in the central region $|y|<0.8$.

A similar measurement (rapidity gap as a function of $Q_0$ and $Q_{\rm sum}$
unfolded to the particle level) was done by CMS~\cite{CMS-PAS-TOP-12-023}
in the dilepton channel with 5~fb$^{-1}$ of 7~TeV data.
The rapidity gap distributions were determined for the whole jet rapidity range
($|y|<2.4$) and compared to predictions from
{\sc MadGraph}~\cite{Alwall:2011uj} and
{\sc Powheg} interfaced to {\sc Pythia}, and
{\sc MC@NLO} interfaced to {\sc Herwig}.
The best description was obtained with {\sc MC@NLO}, while it was observed
that increasing the $Q^2$ scale in {\sc MadGraph} improves the
agreement between the data and the simulation.

\subsection{Jet multiplicity in \ttbar\ events and \ttbar+jet production cross section}
The first measurement of the cross section of \ttbar\ with an additional
jet was performed at CDF using 4.1~fb$^{-1}$ of integrated luminosity~\cite{CDF-ttj}. The measurement was performed in the lepton+jets channel,
with a 2D likelihood discriminant formed to simultaneously measure
the \ttbar+jet and \ttbar+(no jet) cross sections. The result
$\sigma(\ttbar+j)=1.6\pm 0.2{\rm(stat.)}\pm 0.5{\rm(syst.)}$~pb
was found to be in a good agreement with next-to-leading order QCD
calculations.

The jet multiplicity distribution in \ttbar\ events was measured
by ATLAS in the lepton+jet channel
and by CMS in the lepton+jet and dilepton channels.
Both sets of results were obtained at a center-of-mass energy of 7~TeV,
using 4.7~fb$^{-1}$ (ATLAS) and 5~fb$^{-1}$ (CMS) of integrated
luminosity. ATLAS measured~\cite{ATLAS-CONF-2012-155}
the jet multiplicity for four values of the
jet transverse momentum threshold (25, 40, 60, and 80~GeV).
The results were unfolded to the particle level with a response
matrix $M_{\rm part}^{\rm reco}$ applied iteratively using Bayesian
unfolding~\cite{D'Agostini:1994zf}. The resulting
jet multiplicity distributions were found to be consistent with
{\sc Alpgen} interfaced with {\sc Pythia} and {\sc Herwig},
as well as {\sc Powheg} interfaced with {\sc Pythia}.
As expected, the data disfavours the {\sc MC@NLO} model, which predicts
lower jet multiplicity and softer jets.

CMS measured~\cite{CMS-PAS-TOP-12-023}
jet multiplicity in the dilepton channel for two jet
transverse momentum thresholds (30 and 60~GeV). The results were
corrected to the particle level using a regularised unfolding
method~\cite{Hocker:1995kb,Blobel:2002pu}.
The resulting distributions were compared to Monte Carlo predictions.
As in the ATLAS analysis, it was found that
{\sc MC@NLO} interfaced to {\sc Herwig} generates lower
multiplicities than observed.
{\sc MadGraph} and {\sc Powheg} interfaced to {\sc Pythia}
describe the data with up to six additional jets well.

In the lepton+jet analysis~\cite{CMS-PAS-TOP-12-018}, CMS
measured normalized differential cross section defined as the
measured cross section in each jet multiplicity bin
divided by the measured total cross section in the same phase space.
The results were compared to predictions from various Monte Carlo generators
with conclusions similar to those from the dilepton analysis.
%Normalized cross sections were found to be
%\begin{equation}
%\begin{array}{llll}
%\sigma(\ttbar+0{\rm~jets})    & 0.578 & \pm 0.015 {(\rm stat.)} & \pm 0.028 {(\rm syst.)} \\
%\sigma(\ttbar+1{\rm~jets})    & 0.292 & \pm 0.010 {(\rm stat.)} & \pm 0.029 {(\rm syst.)} \\
%\sigma(\ttbar+\ge2{\rm~jets}) & 0.130 & \pm 0.004 {(\rm stat.)} & \pm 0.019 {(\rm syst.)} \\
%\end{array}
%\end{equation}

ATLAS performed~\cite{ATLAS-CONF-2012-083} a measurement of
the cross section for production of \ttbar\ with additional jets
(denoted $\ttbar j$)
in the lepton+jets channel with 4.7~fb$^{-1}$ of 7~TeV data.
Two definitions of $\ttbar j$ events were considered,
both based on particle jets from the Monte Carlo model.
In definition 1, events were declared $\ttbar j$ if
at least one particle jet could not be matched to a parton from a top quark.
In definition 2, all events with at least five particle jets were
termed $\ttbar j$. The $\ttbar j$ production cross sections were
obtained by matching the Monte Carlo predictions
for $\ttbar j$ and non-$\ttbar j$ events against the data.
The cross section for \ttbar\ production in association with
at least one additional jet according to definition 1 is measured to be
$\sigma_{\ttbar j}=102\pm 2{\rm(stat.)}^{+23}_{-25}{\rm(syst.)}$~pb, and
its ratio to the \ttbar\ inclusive cross section is
$\sigma_{\ttbar j}/\sigma_{\ttbar}^{\rm~incl}=0.54\pm 0.01{\rm(stat.)}^{+0.05}_{-0.08}{\rm(syst.)}$.

\subsection{\ttbar+jets heavy flavor composition}
A first measurement of the cross section ratio
$\sigma(\ttbar\bbbar)/\sigma(\ttbar jj)$ was done by
CMS~\cite{CMS-PAS-TOP-12-024} in the dilepton channel
using 5~fb$^{-1}$ of integrated luminosity collected at
a center-of-mass-energy of 7~TeV. The result was obtained by
fitting the $b$-tagged jet multiplicity distributions and
correcting to particle level. The resulting ratio was found to be
$\sigma(\ttbar\bbbar)/\sigma(\ttbar jj)=3.6\pm 1.1{\rm(stat.)}\pm 0.9{\rm(syst.)}$\% which can be compared to predictions from {\sc MadGraph} (1.2\%) and
{\sc Powheg} (1.3\%). The result cannot be directly compared to
NLO QCD calculations as the particle-to-parton correction would be required.